%% file: paper.tex
\documentclass[%
preprint,
 amsmath,amssymb,
 aps, 
 prresearch,
 longbibliography,
]{revtex4-2}

\usepackage{graphicx}
\usepackage{dcolumn}
\usepackage{bm}
\usepackage[colorlinks=true, allcolors=blue]{hyperref}
\usepackage{verbatim}

\usepackage{siunitx}
\DeclareSIUnit\atm{atm}
\DeclareSIUnit\dyn{dyn}

\newcommand*\chem[1]{\ensuremath{\mathrm{#1}}}

\usepackage{xcolor}

\usepackage{pdfpages}

\makeatletter
\AtBeginDocument{\let\LS@rot\@undefined}
\makeatother

\begin{document}


\title{Giant Nonequilibrium Fluctuations at a Reactive Surface}

\author{Hyun Tae Jung}
\affiliation{Department of Chemistry, Korea Advanced Institute of Science and Technology, Daejeon 34141, South Korea}

\author{Hyungjun Kim}
\email{linus16@kaist.ac.kr}
\affiliation{Department of Chemistry, Korea Advanced Institute of Science and Technology, Daejeon 34141, South Korea}

\author{Alejandro L. Garcia}
\affiliation{Department of Physics and Astronomy, San Jose State University, San Jose, California 95192, USA}

\author{Andrew J. Nonaka}
\affiliation{Center for Computational Sciences and Engineering, Lawrence Berkeley National Laboratory, Berkeley, California 94720, USA}

\author{John B. Bell}
\affiliation{Center for Computational Sciences and Engineering, Lawrence Berkeley National Laboratory, Berkeley, California 94720, USA}

\author{Ishan Srivastava}
\affiliation{Center for Computational Sciences and Engineering, Lawrence Berkeley National Laboratory, Berkeley, California 94720, USA}

\author{Changho Kim}
\email{ckim103@ucmerced.edu}
\affiliation{Department of Applied Mathematics, University of California, Merced, California 95343, USA}

\date{\today}

\begin{abstract}
We investigate whether giant fluctuations in a gas can induce corresponding fluctuations on a reactive surface in contact with the gas. Numerical simulations of a minimal heterogeneous catalytic reactor demonstrate that such fluctuations indeed emerge on the surface, with spatial correlations extending over micrometer scales. These fluctuations originate from the dependence of the adsorption rate on the reactant partial pressure. As a result, the surface-coverage structure factor mirrors that of the partial pressure, exhibiting similar enhancement and roll-off behavior across wave numbers.
\end{abstract}

\maketitle

\section{Introduction}

Giant nonequilibrium fluctuations, originally observed in pioneering experimental observations by Vailati and Giglio~\cite{VailatiGiglio1997}, and Brogioli \textit{et al.}~\cite{BrogioliVailatiGiglio2000}, are now recognized as a universal feature of fluids held out of equilibrium by macroscopic concentration gradients.
Their universality originates from a generic physical mechanism whereby thermal velocity fluctuations couple to macroscopic gradients, giving rise to long-lived, long-ranged correlations of fluctuations characterized by a power-law divergence of the static structure factor, $S(q)\sim q^{-4}$, at intermediate wave vectors~\cite{VailatiGiglio1998}.
At the largest scales, this divergence is suppressed by stabilizing mechanisms such as gravity or confinement, which introduce well-defined cutoffs~\cite{ZarateSengers2006}.
Within this framework, early theoretical descriptions focused on weakly nonequilibrium bulk fluids, where local equilibrium holds and the hydrodynamic equations can be linearized.
This led to the establishment of fluctuating hydrodynamics~\cite{LandauLifschitz1987, GarciaBellNonakaSrivastavaLadigesKim2026} as a standard and predictive description of giant nonequilibrium fluctuations.

Building on this well-established framework, subsequent theoretical, numerical, and experimental studies have extended the investigation of giant nonequilibrium fluctuations to increasingly realistic bulk systems.
These efforts have encompassed a broad range of research directions, including stronger driving and transient nonequilibrium gradients, complex and multicomponent fluids, and geometries with confinement or reduced dimensionality~\cite{DonevBellDeLaFuenteGarcia2011, TakacsVailatiCerbinoMazzoniGiglioCannell2011, ZarateKirkpatrickSengers2015, CroccoloZarateSengers2016, VailatiBatallerBouAliEtAl2023}. 
Together, these studies for both liquids and gases demonstrate the robustness of giant nonequilibrium fluctuations across diverse fluid systems under a wide range of nonequilibrium conditions.
Despite this progress, extensions to fluids subject to chemical reactions have remained comparatively limited~\cite{KimNonakaBellGarciaDonev2018}. 
In particular, neither the existence nor the characteristics of giant nonequilibrium fluctuations at reactive fluid--solid interfaces have been established. 
In this work, we use computer simulations to demonstrate giant nonequilibrium fluctuations at a reactive surface in direct contact with a fluid sustaining a macroscopic gradient, and to characterize their fundamental properties.

\section{\label{sec:Model}Reactive Gas--Solid Interface Model}

\begin{figure}
    \includegraphics[width=0.6\columnwidth]{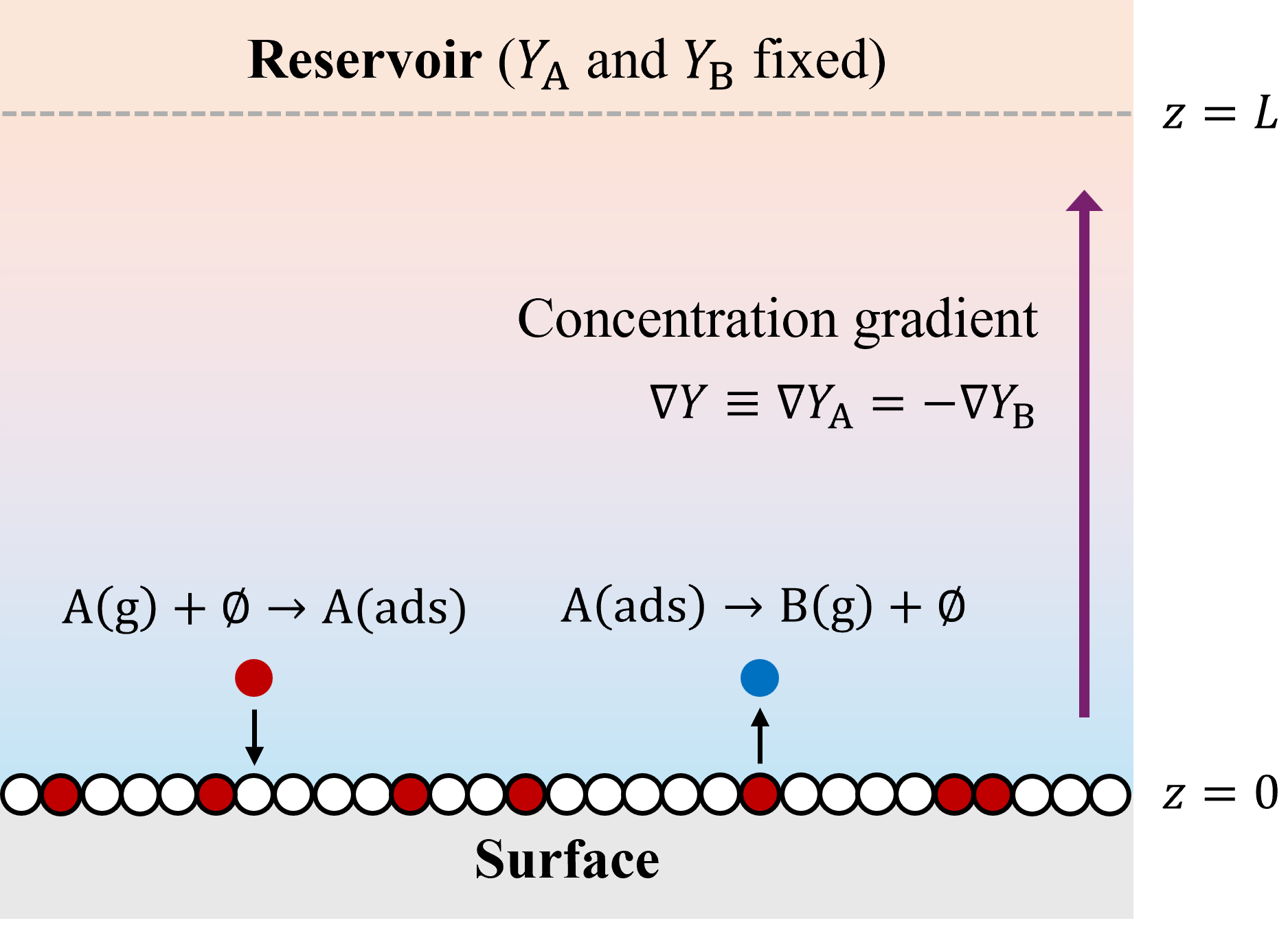}
    \caption{\label{Fig:System}
    Schematic illustration of a gas--solid interfacial system in which the conversion of \chem{A} (red) to \chem{B} (blue) occurs at a reactive surface.
    The reservoir fixes the concentrations of \chem{A} and \chem{B} so the surface reaction induces concentration gradients normal to the surface for both species.}
\end{figure}

We consider a gas--solid interfacial system that can be viewed as a simple heterogeneous catalytic reactor (see Fig.~\ref{Fig:System}). 
The gas phase consists of an ideal mixture of species~\chem{A} and \chem{B}, while the reactive surface, located at $z=0$, supports the following reactions:
\begin{subequations}\label{eq:reactions}
\begin{align}
    &\chem{A}(\mathrm{g}) + \varnothing \xrightarrow{k_a} \chem{A}(\mathrm{ads}),\\
    &\chem{A}(\mathrm{ads}) \xrightarrow{k_d} \chem{B}(\mathrm{g}) + \varnothing,
\end{align}
\end{subequations}
where $\varnothing$ denotes an empty adsorption site, and $k_a$ and $k_d$ are the rate constants for adsorption and desorption, respectively.
In addition, a reservoir fixes the concentrations of species~\chem{A} and \chem{B} at $z=L$.
As a consequence of the overall surface reaction $\chem{A}\rightarrow\chem{B}$, a linear concentration gradient normal to the surface develops for both species in the steady state.
The central question addressed in this work is whether giant fluctuations arise in the dynamics of the surface coverage, denoted by $\theta$, in the presence of concentration gradients in the contacting gas.

To focus on the essential physics, we introduce the following simplifying assumptions.
First, we assume that the adsorption of the reactant (\chem{A}) and the desorption of the product (\chem{B}) obey the Langmuir adsorption model, and that the reactive surface is in contact with an infinite heat bath so that it is held at a constant temperature $T$~\cite{JungKimGarciaNonakaBellSrivastavaKim2026}.
We further assume that the temperature in the reservoir~\cite{SrivastavaLadigesNonakaGarciaBell2023} is also held fixed at $T$.
Finally, we assume that species~\chem{A} and \chem{B} have identical molecular properties and that there are no heats of reaction for \eqref{eq:reactions}.
As a result, if these species were not distinguished, the system would be in thermal equilibrium at temperature $T$.
However, we treat them as distinguishable and use the mass fraction $Y \equiv Y_\chem{A} = 1 - Y_\chem{B}$ as the measure of concentration.
The steady state is determined by the balance of adsorption, desorption, and diffusive fluxes at the gas--solid interface:
\begin{equation}
\label{eq:steady state}
    k_a\: \bar{p}_\mathrm{tot} \bar{Y}^{(\mathrm{surf})} (1-\bar{\theta}) 
    = k_d\: \bar{\theta}
    = \frac{\bar{\rho}_\mathrm{tot} D\: \lvert \overline{\nabla Y} \rvert}{m\: n_\mathrm{site}}.
\end{equation}
Here, the overbar denotes a macroscopic mean; $p_\mathrm{tot}$ and $\rho_\mathrm{tot}$ are the total pressure and total mass density of the ideal gas mixture, respectively; $Y^{(\mathrm{surf})}$ is the mass fraction of species~\chem{A} immediately above the surface; $\lvert \overline{\nabla Y} \rvert = (Y^{(\mathrm{res})}-\bar{Y}^{(\mathrm{surf})})/L$ is the linear concentration gradient of species~\chem{A}, where $Y^{(\mathrm{res})}$ is the fixed mass fraction of \chem{A} in the reservoir; $D$ is the binary diffusion coefficient; $m$ is the mass of the gas molecules; and $n_\mathrm{site}$ is the surface density of adsorption sites.
Note that $\bar{\theta}$ and $\bar{Y}^{(\mathrm{surf})}$ are given by Eq.~\eqref{eq:steady state} once the various system parameters are specified.

\section{Simulation Method}

We perform three-dimensional computer simulations of the system using a stochastic simulation method that we have recently developed for reactive gas--solid interfacial systems~\cite{JungKimGarciaNonakaBellSrivastavaKim2026}.
In this method, the gas-phase dynamics is simulated using a computational fluctuating hydrodynamics approach~\cite{SrivastavaLadigesNonakaGarciaBell2023}, while the surface coverage dynamics is described by a mean-field model that only considers $\theta$ within each computational cell rather than the actual set of sites that are occupied.
This mean-field description is appropriate because the surface reactions considered here involve only single-site processes.

We consider an ideal gas mixture at $T = \SI{273}{\K}$ and $\bar{p}_\mathrm{tot} = \SI{2.03e6}{\dyn\:\cm^{-2}}$.
As a representative small gas molecule, we use molecular properties of \chem{CO}: $m = \SI{4.65e-23}{\g}$, $D = \SI{7.92e-2}{\cm^2\:\s^{-1}}$; for the calculation of the other transport coefficients and the internal energy, we use a molecular diameter of $\SI{3.76e-8}{\cm}$ and a constant-volume specific heat capacity of $\frac{5}{2}k_B/m$.
Parameter values relevant to the gas--solid interface are summarized in Table~\ref{Tab:parameters}.
The computational domain is a rectangular box of size $8L \times L \times L$, with periodic boundary conditions imposed in the $x$ and $y$ directions.
The domain is spatially discretized into $1024\times128\times128$ uniform cubic cells with grid spacing $\Delta x=\SI{2.81e-5}{\cm}$.
Each bottom-layer cell has a face in contact with the surface of area $\Delta x^2$ containing $900\times900$ adsorption sites, corresponding to a surface site density $n_\mathrm{site}=\SI{1.03e15}{\cm^{-2}}$.

\begin{table}
    \caption{\label{Tab:parameters}
        Physical parameter values used in the three-dimensional simulations.
        These serve as a baseline parameter set for the two-dimensional simulations, in which selected parameters are varied to systematically explore their effects.}
    \begin{tabular}{|c|c||c|c|}
        \hline
        $k_a$ & \SI{29.3}{\cm^2\:\dyn^{-1}\:\s^{-1}} & $k_d$ & \SI{1.58e+06}{\s^{-1}} \\
        \hline
        $\bar{Y}^{(\mathrm{surf})}$ & \SI{4.46e-02}{} & $Y^{(\mathrm{res})}$ & 0.900 \\
        \hline
        $\bar{\theta}$ & 0.626 & $L$ & \SI{3.59e-03}{\cm} \\
        \hline
    \end{tabular}
\end{table}

In this work, we numerically compute two static structure factor spectra, $S_\theta(\mathbf{k}_\perp)$ and $S_Y^{(\mathrm{surf})}(\mathbf{k}_\perp)$, characterizing the spatial correlations of fluctuations occurring at or near the reactive surface.
These spectra are obtained from lateral Fourier transforms over cross sections of the system normal to the $z$ axis.
For the three-dimensional system, $\mathbf{k}_\perp = (k_x,k_y)$ and the structure factor results are presented as a function of $k_\perp=\lvert \mathbf{k}_\perp\rvert$.
To compute $S_Y^{(\mathrm{surf})}$, we use the values of $Y$ in the bottom-layer cells at the reactive surface and normalize by the cell volume $\Delta V= \Delta x^3$, i.e.,  $S_Y^{(\mathrm{surf})}(k_\perp) = \Delta V \langle \delta\hat{Y}^{(\mathrm{surf})} (\delta\hat{Y}^{(\mathrm{surf})})^*\rangle$.
For $S_\theta$, we normalize by unity, i.e., $S_\theta(k_\perp)= \langle \delta\hat{\theta}(\delta\hat{\theta})^*\rangle$, where $\delta\hat{\theta}$ is the Fourier transform of $\delta\theta = \theta - \bar{\theta}$.
We employ unity normalization for the forward discrete Fourier transform.
To account for the effect of the discrete Laplacian, we use a modified wave vector~\cite{KimNonakaBellGarciaDonev2017}; however, this effect is negligible in the small-wave-number regime of interest in this work.
For all simulations, the structure factors are computed over $10^7$ time steps after reaching steady state, where the time step is $\Delta t=\SI{1e-10}{s}$.
The equilibrium structure factors of the surface coverage and mass fraction are given by~\cite{JungKimGarciaNonakaBellSrivastavaKim2026}
\begin{equation}
    S_{\theta,\mathrm{eq}} = \frac{\bar{\theta}(1-\bar{\theta})}{N_\mathrm{tot}},\;\;
    S_{Y,\mathrm{eq}}^{(\mathrm{surf})} = \frac{m}{\bar{\rho}_\mathrm{tot}} \bar{Y}^{(\mathrm{surf})} [1-\bar{Y}^{(\mathrm{surf})}].
\end{equation}

\section{Results and Discussion}

\subsection{Giant Fluctuations at a Reactive Surface}

\begin{figure*}
    \includegraphics[width=\linewidth]{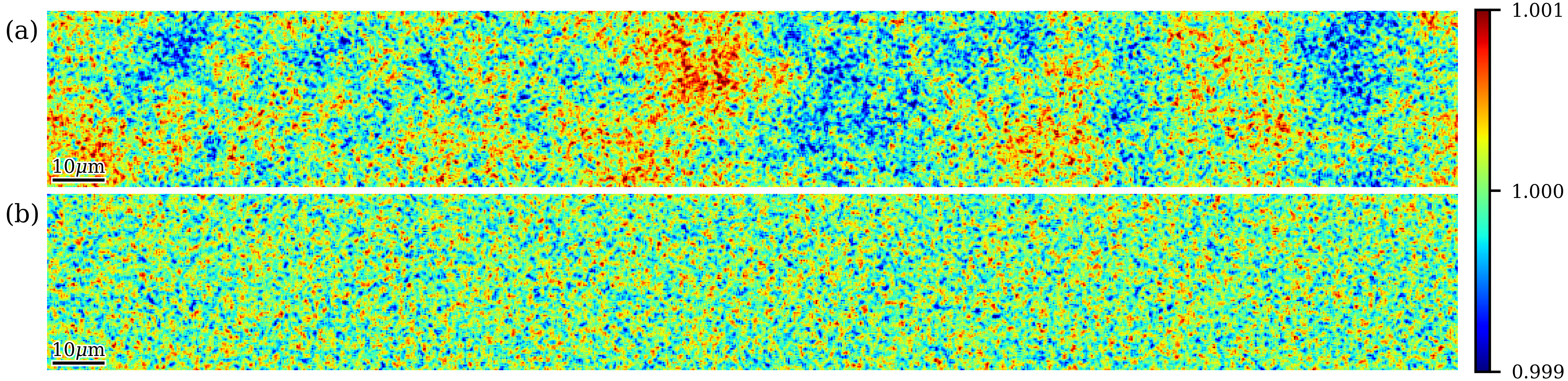}
    \caption{\label{Fig:Snapshot}
        Snapshots of the normalized surface coverage, $\theta(x,y)/\bar{\theta}$, over the entire domain $8L\times L$.
        Panel~(a) shows the nonequilibrium simulation, whereas panel~(b) shows the corresponding equilibrium simulation.
        In both cases, a slip boundary condition for velocity is imposed at the surface and the surface coverage is boxcar-averaged using a $3\times3$ window on a $1024\times128$ grid.}
\end{figure*}

In Fig.~\ref{Fig:Snapshot}, we demonstrate the emergence of long-ranged fluctuations in the surface coverage.
This is illustrated by comparing a snapshot of the surface coverage over the entire surface (size $8L\times L$) at a statistical steady state with a comparable equilibrium system, in which the conversion reaction is replaced by reversible adsorption of \chem{A}, $\chem{A}(\mathrm{g}) + \varnothing \rightleftharpoons \chem{A}(\mathrm{ads})$, with $Y^{(\mathrm{res})} = \bar{Y}^{(\mathrm{surf})}$.
The nonequilibrium simulation exhibits remarkably long-ranged spatial correlations, with fluctuations organized into extended structures whose correlation length exceeds several micrometers (the system width in Fig.~\ref{Fig:Snapshot} is about \SI{0.3}{\mm}).
By contrast, the equilibrium snapshot shows no spatial organization: fluctuations are short-ranged and uncorrelated, producing a featureless, noise-like pattern.
This behavior in the nonequilibrium case is not attributable to typically short-ranged surface-chemical lateral interactions.
Instead, it arises from giant fluctuations in the partial pressure (or mass fraction) of species~\chem{A} in the contacting gas, which induces long-ranged correlations in the adsorption rate.

\begin{figure}
    \includegraphics[width=0.6\columnwidth]{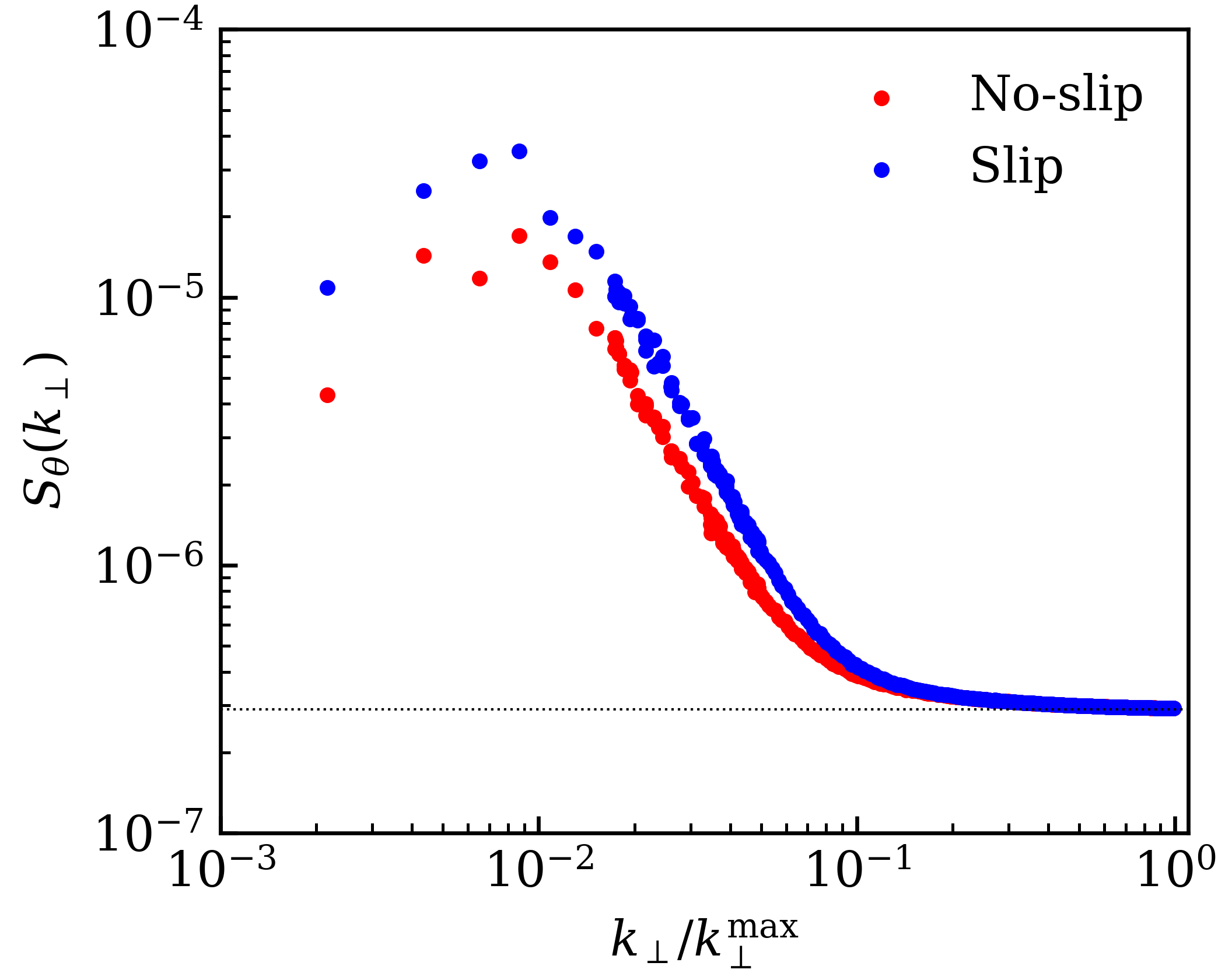}
    \caption{\label{Fig:3D Structure Factor}
        Structure factor spectra of the surface coverage $S_\theta(k_{\perp})$ obtained from simulations with slip and no-slip velocity boundary conditions at the reactive surface.
        The lateral wave number $k_{\perp}$ is normalized by its maximum value, $k_{\perp}^{\mathrm{max}} = 2\sqrt{2}/\Delta x$.
        The dotted line denotes the corresponding equilibrium structure factor, $S_{\theta,\mathrm{eq}}$.
        For all structure factor spectra presented in this work (i.e., Figs.~\ref{Fig:3D Structure Factor}--\ref{Fig:FiniteSizeEffect}), bin averaging is applied in the large-wave-number regime ($k_\perp/k_\perp^\mathrm{max} \ge 0.05$) to reduce the number of data points for clarity.}
\end{figure}

To analyze these long-ranged correlations quantitatively, we compute the structure factor of the surface coverage, $S_\theta(k_\perp)$.
Fig.~\ref{Fig:3D Structure Factor} shows that the surface-coverage structure factor approaches its equilibrium value at large wave numbers, but as $k_\perp$ decreases it displays a pronounced enhancement---reaching amplitudes up to two orders of magnitude larger than the equilibrium value---followed by a roll-off at the smallest wave numbers.
This qualitative behavior is observed for both no-slip and slip boundary conditions for velocity at the surface.
An important observation is that $S_\theta(k_\perp)$ closely resembles the spectral features associated with giant nonequilibrium fluctuations in confined fluid systems (see Fig.~S1 in the Supplemental Material~\cite{SM}).

\subsection{Probing the Origin of Surface Fluctuations}

To investigate the connection between the surface coverage structure factor, $S_\theta(k_\perp)$, and the structure factor of the mass fraction immediately above the surface, $S_Y^{(\mathrm{surf})}(k_\perp)$, we perform quasi-two-dimensional simulations over a range of parameter values.
Here, the quasi-two-dimensional system refers to an $xz$-slice of the three-dimensional system (i.e., $8L\times L$ with cell depth $\Delta x$) and $\mathbf{k}_\perp$ reduces to $k_x$.

We use this reduced geometry solely for computational efficiency in exploring parameter space.
As expected, the resulting structure factor spectra show quantitative agreement with the three-dimensional simulations as demonstrated in Fig.~S2 of the Supplemental Material~\cite{SM}.

\begin{figure}
    \includegraphics[width=0.6\columnwidth]{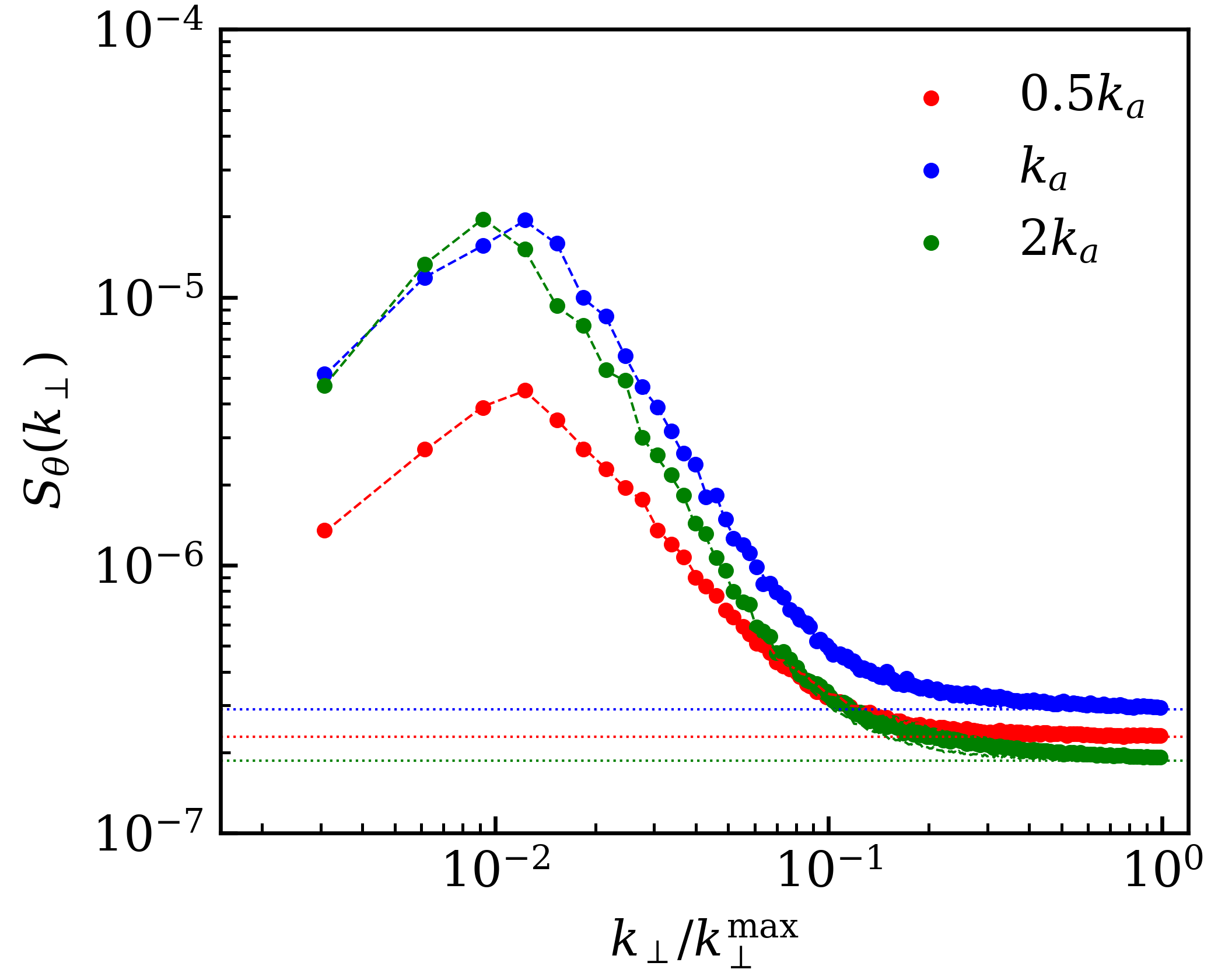}
    \caption{\label{Fig:Linear Relationship}
        Surface coverage structure factor spectra for different adsorption rate constants under the same imposed concentration gradient.
        The results labeled $k_a$ (blue) are obtained from a quasi-two-dimensional simulation with the parameter values listed in Table~\ref{Tab:parameters}.  
        For the cases labeled $0.5k_a$ and $2k_a$ (red and green, respectively), the adsorption rate constant is varied while all other parameters, except $\bar{\theta}$ and $k_d$, are held fixed.
        The corresponding values of $\bar{\theta}$ and $k_d$ are adjusted to satisfy Eq.~\eqref{eq:steady state}.
        The dashed lines show $S_\theta$ computed from Eq.~\eqref{eq:Computed SFtheta} using the corresponding numerical results for $S_Y^{(\mathrm{surf})}$, while the dotted lines denote $S_{\theta,\mathrm{eq}}$.
        For all quasi-two-dimensional results shown in Figs.~\ref{Fig:Linear Relationship}--\ref{Fig:Weak Coupling}, the lateral wave number $k_{\perp}$ is normalized by its maximum value, $k_{\perp}^{\mathrm{max}} = 2/\Delta x$, and a no-slip boundary condition is imposed at the surface.}
\end{figure}

The nonequilibrium enhancement of the surface coverage ($S_{\theta,\mathrm{neq}} = S_\theta - S_{\theta,\mathrm{eq}}$) is expected to be closely related to that of the interfacial mass fraction ($S_{Y,\mathrm{neq}}^{(\mathrm{surf})} = S_Y^{(\mathrm{surf})} - S_{Y,\mathrm{eq}}^{(\mathrm{surf})}$).
This motivates the scaling relation
\begin{equation} 
\label{eq:SFscaling} 
    S_{\theta,\mathrm{neq}} \propto \frac{1}{\Delta V} S_{Y,\mathrm{neq}}^{(\mathrm{surf})}.
\end{equation} 
Here, the factor $1/\Delta V$ arises from the different normalization conventions used for $S_\theta$ and $S_Y^{(\mathrm{surf})}$.
Our numerical results show that, for each set of parameters considered, this proportionality provides a good description with a $k_\perp$-independent proportionality coefficient over the entire range of $k_\perp$ examined.
More specifically, we find that
\begin{equation}
\label{eq:Computed SFtheta}
    S_\theta = \frac{1}{\Delta V} \left[\frac{\bar{\theta}(1-\bar{\theta})}{\bar{Y}^{(\mathrm{surf})}}\right]^2 \left(S_Y^{(\mathrm{surf})} - S_{Y,\mathrm{eq}}^{(\mathrm{surf})}\right) + S_{\theta,\mathrm{eq}}.
\end{equation}
In Fig.~\ref{Fig:Linear Relationship}, we compare $S_\theta$ obtained directly with the values computed from Eq.~\eqref{eq:Computed SFtheta} using $S_Y^{(\mathrm{surf})}$ for various values of $k_a$. 
The agreement is excellent at small $k_\perp$ and remains good over the entire range of $k_\perp$ examined.
The wave-number dependence of this agreement is further quantified in Fig.~S3 of the Supplemental Material~\cite{SM}.
The same level of agreement is also found when a slip velocity boundary condition is imposed at the surface, as shown in Fig.~S4 of the Supplemental Material~\cite{SM}.

The origin of the proportionality factor in Eq.~\eqref{eq:Computed SFtheta}, $\left[\bar{\theta}(1-\bar{\theta})/\bar{Y}^{(\mathrm{surf})}\right]^2$, can be understood from the linearized stochastic equation for $\delta\theta$ about the steady state:
\begin{equation}
\label{eq:dtheta}
    d(\delta\theta) = 
    \bar{r} \left( 
        \frac{\delta Y^{(\mathrm{surf})}}{\bar{Y}^{(\mathrm{surf})}} 
        - \frac{\delta\theta}{\bar{\theta}(1-\bar{\theta})} 
    \right) dt + \sqrt{\frac{2\bar{r}}{N_\mathrm{tot}}} dW
\end{equation}
where $\bar{r} \equiv k_a\: \bar{p}_\mathrm{tot} \bar{Y}^{(\mathrm{surf})} (1-\bar{\theta}) = k_d\bar{\theta}$ is the mean reaction rate (see Eq.~\eqref{eq:steady state}), $N_\mathrm{tot} = n_\mathrm{site}\Delta x^2$ is the total number of adsorption sites per cell, and $W$ denotes a standard Wiener process.
A detailed derivation of Eq.~\eqref{eq:dtheta}, including the stochastic notation, is provided in Appendix~\ref{sec:appendixLineardthetaDerivation}.

As shown in Appendix~\ref{sec:appendixStructureFactor}, in the small-$k_\perp$ regime, the mass-fraction fluctuations vary slowly compared with the relaxation of the surface coverage, allowing $\delta\theta$ to respond quasi-statically to $\delta Y^{(\mathrm{surf})}$. 
This time-scale separation provides a theoretical motivation for the proportionality factor in Eq.~\eqref{eq:Computed SFtheta}.
However, a complete analytical description of the structure-factor relation in Eq.~\eqref{eq:Computed SFtheta} over the full range of $k_\perp$ would require a structure-factor analysis of the coupled gas--solid interfacial system.
While such an analysis has recently been carried out for the equilibrium case~\cite{JungKimGarciaNonakaBellSrivastavaKim2026}, a corresponding treatment of the present nonequilibrium system will be the subject of future work.

\begin{figure}
    \includegraphics[width=0.6\columnwidth]{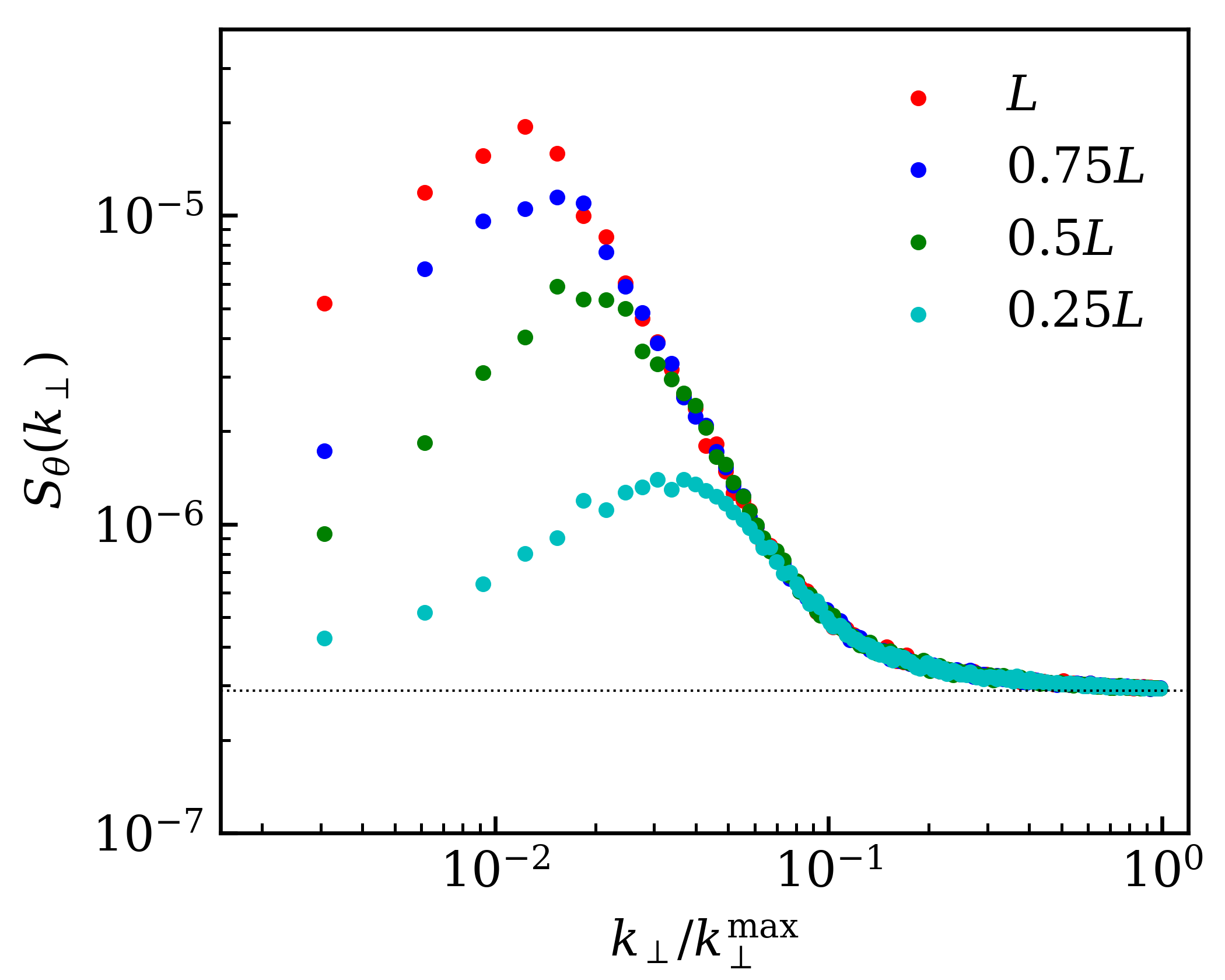}
    \caption{\label{Fig:FiniteSizeEffect}
        Surface coverage structure factor spectra for different system confinement lengths in the gradient direction under the same imposed concentration gradient.
        The results labeled $L$ (red) are obtained from a quasi-two-dimensional simulation with the parameter values listed in Table~\ref{Tab:parameters}.
        For the other cases, the system confinement length is varied while all other parameters are held fixed, except for $\bar{Y}^{(\mathrm{res})}$, which is adjusted to obtain the same imposed concentration gradient.
        The dotted line denotes the corresponding equilibrium structure factor, $S_{\theta,\mathrm{eq}}$.}
\end{figure}

We examine the effect of confinement in the gradient direction on the nonequilibrium enhancement of the surface coverage structure factor $S_\theta$.
To this end, we compute surface coverage structure factor spectra for several confinement lengths of the system while keeping the imposed concentration gradient fixed.
As shown in Fig.~\ref{Fig:FiniteSizeEffect}, decreasing the confinement length shifts the roll-off to larger wave numbers and leads to a stronger suppression of the nonequilibrium enhancement.
Correspondingly, the wave number $k_\perp^\mathrm{peak}$ at which $S_\theta$ attains its maximum is observed to be inversely proportional to the confinement length.
These behaviors can be qualitatively understood by invoking the linear relation between $S_\theta$ and $S_Y^{(\mathrm{surf})}$ [Eq.~\eqref{eq:Computed SFtheta}], provided that $S_Y^{(\mathrm{surf})}$ in the present reactive system exhibits a confinement-induced roll-off similar to that observed in nonreactive confined bulk fluid systems~\cite{ZarateRedondo2001}.
In this sense, the finite-size suppression of giant nonequilibrium fluctuations in the gas phase is transmitted to the surface coverage through the interfacial coupling.

\subsection{Passive versus Active Interfacial Coupling}

The qualitative interpretation of the confinement effects discussed above assumes that $S_\theta$ reflects $S_Y^{(\mathrm{surf})}$ because the surface chemistry passively transmits gas-phase fluctuations without significantly modifying $S_Y^{(\mathrm{surf})}$ itself.
To assess the validity of this passive-coupling interpretation, we consider a weak-coupling limit in which the surface chemistry induces a concentration gradient in the gas phase but does not otherwise modify the giant nonequilibrium fluctuations of the gas. 
In this limit, the enhancement of $S_Y^{(\mathrm{surf})}$ is identical to that of a hypothetical nonreactive system with the same gradient, and the giant fluctuations in $S_\theta$ are passively induced through the linear relation given by Eq.~\eqref{eq:Computed SFtheta}. 
For various values of the surface-reaction relaxation time $\tau_\theta = \bar{\theta}(1-\bar{\theta})/\bar{r}$ (see Eq.~\eqref{eq:dtheta}), 
we compute the weak-coupling predictions for $S_\theta$ and compare them with our simulation results.
We estimate $S_Y^{(\mathrm{surf})}$ for the hypothetical nonreactive system using Eq.~\eqref{eq:nonreactiveSYsurf}; 
details of this hypothetical system and the approximation for $S_Y^{(\mathrm{surf})}$ are provided in Appendix~\ref{sec:appendixNonreactive}.

\begin{figure}
    \includegraphics[width=0.6\columnwidth]{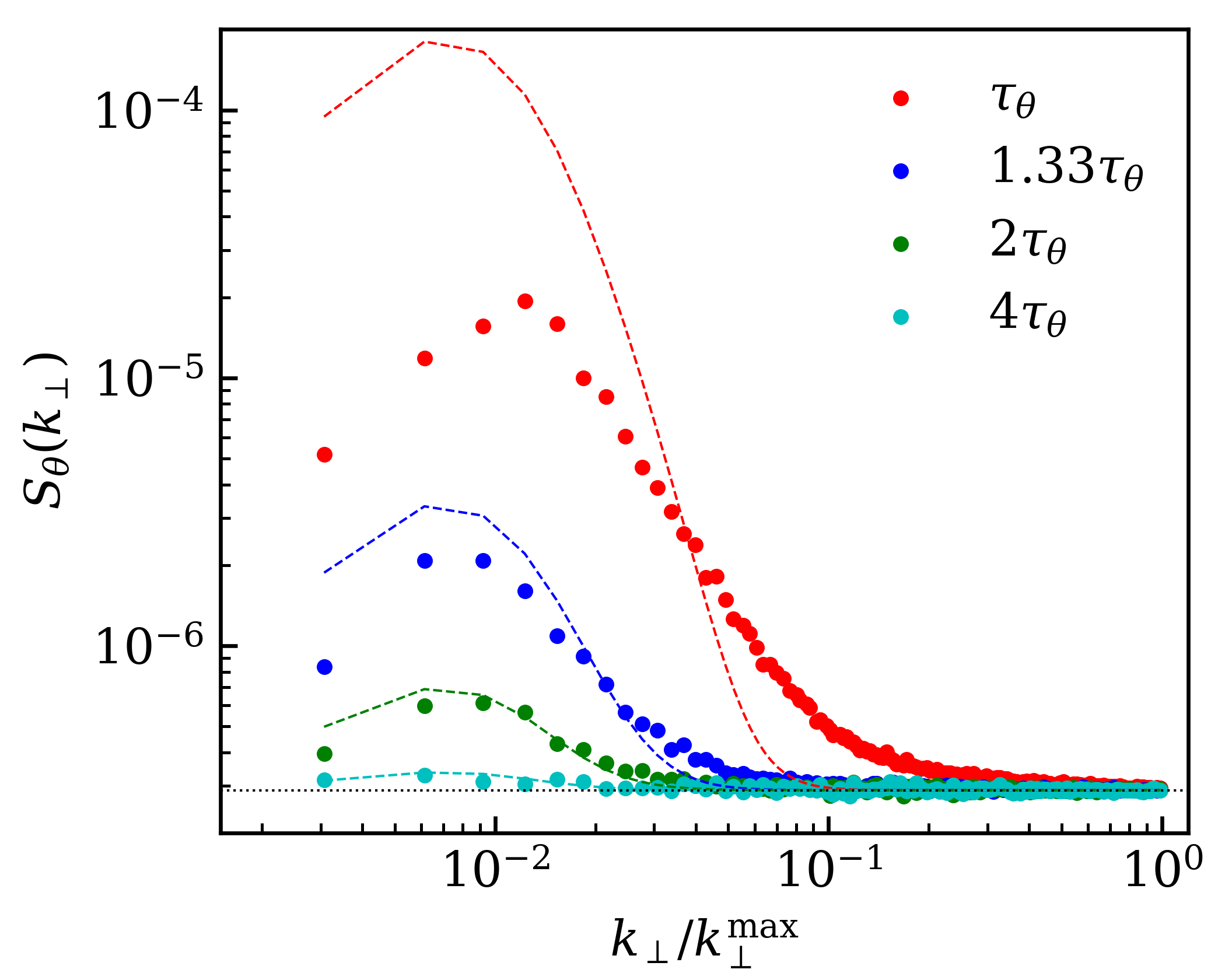}
    \caption{\label{Fig:Weak Coupling}
        Surface coverage structure factor spectra for different surface-reaction relaxation times at the same imposed mean surface coverage.
        The results labeled $\tau_\theta$ (red) are obtained from a quasi-two-dimensional simulation with the parameter values listed in Table~\ref{Tab:parameters}.
        For the other cases, the values of $k_d$, $k_a$, and $\bar{Y}^{(\mathrm{surf})}$ are chosen to produce the corresponding relaxation times while satisfying Eq.~\eqref{eq:steady state}.
        The dashed lines show the weak-coupling predictions for $S_\theta$, computed from Eq.~{\eqref{eq:Computed SFtheta}} using $S_Y^{(\mathrm{surf})}$ for the hypothetical nonreactive system, while the dotted line denotes the corresponding equilibrium structure factor, $S_{\theta,\mathrm{eq}}$.
        }
\end{figure}

Fig.~\ref{Fig:Weak Coupling} compares the weak-coupling predictions with the directly computed $S_\theta$ for various values of $\tau_\theta$.
Increasing $\tau_\theta$ leads to a reduction in the overall magnitude of the nonequilibrium enhancement of $S_\theta$.
For larger $\tau_\theta$, the weak-coupling approximation reproduces the enhancement of $S_\theta$ over a broad range of wave numbers. 
As $\tau_\theta$ decreases, however, deviations from the weak-coupling predictions become increasingly pronounced at small wave numbers and extend toward larger wave numbers. 
This behavior indicates that the extent to which the surface acts as a passive probe depends on the surface-reaction conditions as well as on the wave number of the fluctuations.

To characterize this wave-number-dependent crossover between passive and active interfacial coupling, we first identify a characteristic length scale by comparing the time scale over which the surface depletes species $\chem{A}$ from the adjacent gas with that for diffusive replenishment.
For a gas layer of thickness $h$ above the surface, the characteristic depletion time can be estimated as the amount of species $\chem{A}$ in the layer per unit area divided by its mean consumption flux at the surface: 
\begin{equation} 
    \tau_\mathrm{dep}(h) \sim 
    \frac{\bar{\rho}_\mathrm{tot}\bar{Y}^{(\mathrm{surf})}h} {\bar{\rho}_\mathrm{tot}D\lvert\overline{\nabla Y}\rvert} 
    = \frac{\bar{Y}^{(\mathrm{surf})}} {\lvert\overline{\nabla Y}\rvert}\frac{h}{D},
\end{equation}
where $\bar{\rho}_\mathrm{tot}D\lvert\overline{\nabla Y}\rvert = m\,n_\mathrm{site}\,\bar{\theta}(1-\bar{\theta})/\tau_\theta$ is the mean mass flux of species $\chem{A}$ consumed by the surface, as given by Eq.~\eqref{eq:steady state}.
By comparison, the characteristic time for diffusion across the same layer is 
\begin{equation} 
    \tau_\mathrm{diff}(h) \sim \frac{h^2}{D}. 
\end{equation} 
The two time scales become comparable at $h\sim L_\mathrm{dep}$, where 
\begin{equation} 
    L_\mathrm{dep} \equiv \frac{\bar{Y}^{(\mathrm{surf})}} {\lvert\overline{\nabla Y}\rvert}.
\end{equation} 
Thus, $L_\mathrm{dep}$ represents the characteristic distance over which surface consumption competes with diffusive replenishment of species $\chem{A}$.
For $h\ll L_\mathrm{dep}$, diffusion replenishes the layer faster than the surface depletes it, whereas surface consumption becomes important for $h\gtrsim L_\mathrm{dep}$.

Since gas fluctuations with lateral wave number $k_\perp$ probe a vertical length scale of order $h\sim 1/k_\perp$, the surface has little influence on fluctuations with $k_\perp L_\mathrm{dep}\gg 1$, whereas it actively reshapes those with $k_\perp L_\mathrm{dep}\lesssim 1$.
Thus, the passive-to-active crossover is governed by the dimensionless parameter $k_\perp L_\mathrm{dep}$. 
The relaxation time $\tau_\theta$ determines how rapidly the surface coverage follows the gas fluctuations [Eq.~\eqref{eq:dtheta}], while varying $\tau_\theta$ under the present steady-state constraints also changes $L_\mathrm{dep}$ and thereby shifts the crossover wave number.  
As $\tau_\theta$ increases, the surface consumes species $\chem{A}$ at a lower rate, which weakens the concentration gradient and raises $\bar{Y}^{(\mathrm{surf})}$. 
Both effects increase $L_\mathrm{dep}$, thereby shifting the estimated crossover wave number, $k_\perp^\mathrm{cross}\sim 1/L_\mathrm{dep}$, toward smaller values. 
Using $k_\perp L_\mathrm{dep}\sim1$ as an estimate of the crossover, we obtain $k_\perp/k_\perp^{\max} \approx 7.5\times10^{-2}$, $9.7\times10^{-3}$, $3.5\times10^{-3}$, and $1.2\times10^{-3}$ for the four cases, respectively, consistent with the wave-number range over which deviations from the weak-coupling predictions are observed in Fig.~\ref{Fig:Weak Coupling}.
A further analysis shows that the mass-fraction structure factor $S_Y^{(z)}$ at a height $z$ above the reactive surface is better predicted by the hypothetical nonreactive system as $z$ increases (see Fig.~S5 in the Supplemental Material~\cite{SM}).
Taken together, these observations highlight the need for a full structure-factor analysis of the coupled gas--solid interfacial system.

\section{Conclusion}

This work establishes that giant nonequilibrium fluctuations are not confined to fluid phases but can emerge as an intrinsic feature of reactive surfaces in contact with fluids out of equilibrium. 
Our simulation results show that long-ranged spatial correlations in surface coverage arise because interfacial chemistry provides a direct pathway by which macroscopic nonequilibrium fluctuations in the bulk are transmitted---and, when surface feedback becomes significant, actively reshaped---by the surface kinetics.
This identifies reactive interfaces as active participants in nonequilibrium fluctuation phenomena, rather than merely as boundary conditions for bulk transport.
A full analytical structure-factor theory for the coupled gas--solid interfacial system, extending existing treatments of bulk nonequilibrium fluctuations, therefore remains an important direction for future work.
In addition, our findings suggest that giant nonequilibrium fluctuations at reactive interfaces should be experimentally observable, for example through spatially resolved measurements of surface coverage or reaction activity in catalytic systems subject to imposed concentration gradients.

\begin{acknowledgments}
This research was supported in part by the U.S.\ National Science Foundation under Grant No.\ CHE-2213368 and by the National Research Foundation of Korea, funded by the Korean government, under Grant No.\ RS-2024-00435493.
The work of AN, JB, and IS was supported by the U.S.\ Department of Energy, Office of
Science, Office of Advanced Scientific Computing Research, Applied Mathematics Program
under Contract No.\ DE-AC02-05CH11231.
\end{acknowledgments}

\section*{Data availability}
The data that support the findings of this study were generated by numerical simulations.
The source code and parameters used to generate the simulations are publicly available~\cite{FHDeX}.

\appendix

\section{\label{sec:appendixLineardtheta}Linearized Analysis of Surface-Coverage Fluctuations}

In this appendix, we first derive the linearized stochastic equation for $\delta\theta$ in Eq.~\eqref{eq:dtheta}, following the approach developed in Ref.~\cite{JungKimGarciaNonakaBellSrivastavaKim2026}, and then use this equation to provide a theoretical motivation for the proportionality factor underlying the scaling relation in Eq.~\eqref{eq:SFscaling}.

\subsection{\label{sec:appendixLineardthetaDerivation}Derivation of the Linearized Stochastic Equation}

We begin by considering the local surface-coverage dynamics of a gas in contact with a reactive surface containing $N_\mathrm{tot}$ adsorption sites.
The reactant mass fraction in the gas immediately above the surface and the surface coverage fluctuate about their mean values $\bar{Y}^{(\mathrm{surf})}$ and $\bar{\theta}$, respectively.

Over an infinitesimal time interval $[t,t+dt]$, let $dN_a$ and $dN_d$ denote the numbers of adsorption and desorption events, respectively. 
Consistent with the stochastic update of the surface coverage used in our simulations, we treat these events as independent Poisson processes, such that $dN_a$ and $dN_d$, conditioned on the instantaneous state at time $t$, are Poisson distributed with means $\lambda_a\, dt$ and $\lambda_d\, dt$, respectively.
The corresponding instantaneous rates are 
\begin{equation} 
\label{eq:lambdaad} 
    \lambda_a = k_a\: p_\mathrm{tot} Y^{(\mathrm{surf})}(1-\theta)N_\mathrm{tot}, \quad 
    \lambda_d = k_d\: \theta N_\mathrm{tot}. 
\end{equation}

Before linearizing Eq.~\eqref{eq:lambdaad}, we identify the fluctuations relevant to the surface-coverage dynamics. 
As discussed in Sec.~\ref{sec:Model}, the system would be in thermal equilibrium if species $\chem{A}$ and $\chem{B}$ were not distinguished. 
The total mass density and temperature are therefore not driven out of equilibrium, and we neglect their fluctuations, along with the resulting pressure fluctuations. 
Thus, the following analysis focuses on $\delta Y^{(\mathrm{surf})}$ and $\delta\theta$.

In the weak-noise limit, we write
\begin{equation}
    Y^{(\mathrm{surf})} = \bar{Y}^{(\mathrm{surf})}\left(1+\frac{\delta Y^{(\mathrm{surf})}}{\bar{Y}^{(\mathrm{surf})}}\right), \quad
    \theta = \bar{\theta}\left(1+\frac{\delta\theta}{\bar{\theta}}\right).
\end{equation}
As in Sec.~\ref{sec:Model}, the adsorption and desorption rates balance at the mean state, with their common rate per adsorption site denoted by $\bar{r}$:
\begin{equation} 
    \bar{r} = k_a \bar{p}_\mathrm{tot}\bar{Y}^{(\mathrm{surf})}(1-\bar{\theta}) = k_d\bar{\theta}. 
\end{equation}
Substituting these expressions into Eq.~\eqref{eq:lambdaad} and retaining terms to linear order gives
\begin{equation}
    \lambda_a = \bar{r} N_{\mathrm{tot}} \left(1+ \frac{\delta Y^{(\mathrm{surf})}}{\bar{Y}^{(\mathrm{surf})}}- \frac{\delta\theta}{1-\bar{\theta}} \right), \quad
    \lambda_d = \bar{r} N_{\mathrm{tot}} \left(1+ \frac{\delta\theta}{\bar{\theta}} \right).
\end{equation}

Following Ref.~\cite{JungKimGarciaNonakaBellSrivastavaKim2026}, we apply the Gaussian approximation to the Poisson processes to obtain 
\begin{subequations} 
\begin{align} 
    dN_a &= \bar{r} N_{\mathrm{tot}} \left(1+ \frac{\delta Y^{(\mathrm{surf})}}{\bar{Y}^{(\mathrm{surf})}}- \frac{\delta\theta}{1-\bar{\theta}} \right)dt + \sqrt{\bar{r}N_{\mathrm{tot}}}\,dW_a, \\ 
    dN_d &= \bar{r} N_{\mathrm{tot}} \left(1+ \frac{\delta\theta}{\bar{\theta}} \right)dt + \sqrt{\bar{r}N_{\mathrm{tot}}}\,dW_d, 
\end{align} 
\end{subequations} 
where $W_a$ and $W_d$ are independent standard Wiener processes associated with adsorption and desorption, respectively. 
Here, $dW_a = W_a(t+dt)-W_a(t)$ and $dW_d = W_d(t+dt)-W_d(t)$ denote their increments over the time interval $[t,t+dt]$; these increments are independent Gaussian random variables with zero mean and variance $dt$.

The resulting change in the number of occupied sites, $dN=dN_a-dN_d$, is
\begin{equation} 
\label{eq:deltaN} 
    dN = \bar{r} N_{\mathrm{tot}} \left( 
        \frac{\delta Y^{(\mathrm{surf})}}{\bar{Y}^{(\mathrm{surf})}} 
        - \frac{\delta\theta}{\bar{\theta}(1-\bar{\theta})} 
        \right) dt + \sqrt{2\bar{r}N_\mathrm{tot}}\,dW.
\end{equation} 
Because $W_a$ and $W_d$ are independent standard Wiener processes, $(W_a-W_d)/\sqrt{2}$ is also a standard Wiener process, which we denote by $W$.
Accordingly, $dW=(dW_a-dW_d)/\sqrt{2}$.
Since the surface coverage is the fraction of occupied sites, $d(\delta\theta)=dN/N_{\mathrm{tot}}$, and Eq.~\eqref{eq:dtheta} follows directly from Eq.~\eqref{eq:deltaN}.

\subsection{\label{sec:appendixStructureFactor}Theoretical Motivation for the Structure-Factor Relation}

To understand the proportionality factor underlying the scaling relation in Eq.~\eqref{eq:SFscaling}, we examine the long-wavelength response implied by Eq.~\eqref{eq:dtheta}.
For this purpose, we consider the spatial Fourier modes of the fluctuations along the surface, denoted by $\delta Y_{\mathbf{k}_\perp}^{(\mathrm{surf})}$ and $\delta\theta_{\mathbf{k}_\perp}$. 
For the spatially extended system, the Wiener process $W$ in Eq.~\eqref{eq:dtheta} represents the local noise associated with each surface cell of area $\Delta x^2$, with independent processes in different cells.
Taking a discrete Fourier transform over the surface cells therefore gives 
\begin{equation} 
    d(\delta\theta_{\mathbf{k}_\perp}) 
    = \bar{r}\left( \frac{\delta Y_{\mathbf{k}_\perp}^{(\mathrm{surf})}} {\bar{Y}^{(\mathrm{surf})}} 
    - \frac{\delta\theta_{\mathbf{k}_\perp}} {\bar{\theta}(1-\bar{\theta})} \right)dt 
    + \sqrt{\frac{2\bar{r}}{N_{\mathrm{tot}}}}\,dW_{\mathbf{k}_\perp}, 
\end{equation} 
where $dW_{\mathbf{k}_\perp}$ denotes the corresponding Fourier-transformed noise increment.

The stationary solution of this linear stochastic equation can be written as
\begin{equation}
    \delta\theta_{\mathbf{k}_\perp}(t) = I_1(t) + I_2(t),
\end{equation}
where
\begin{align}
    I_1(t) 
    &= \frac{\bar{r}}{\bar{Y}^{(\mathrm{surf})}} 
    \int_{-\infty}^t \exp\left[-\frac{\bar{r}}{\bar{\theta}(1-\bar{\theta})}(t-s)\right]
    \delta Y^{(\mathrm{surf})}_{\mathbf{k}_\perp}(s)\; ds, \\
    I_2(t)
    &= \sqrt{\frac{2\bar{r}}{N_\mathrm{tot}}}
    \int_{-\infty}^t \exp\left[-\frac{\bar{r}}{\bar{\theta}(1-\bar{\theta})}(t-s)\right] 
    dW_{\mathbf{k}_\perp}(s).
\end{align}
The exponential kernel in $I_1$ is characterized by the relaxation time $\tau_\theta = \bar{\theta}(1-\bar{\theta})/\bar{r}$, which is independent of $k_\perp$. 
In contrast, the relaxation time of the long-wavelength mass-fraction fluctuations scales as $k_\perp^{-2}$.
Thus, in the small-$k_\perp$ regime of interest, the mass-fraction fluctuations vary slowly compared with the relaxation of the surface coverage.
Over the time interval contributing appreciably to $I_1$, we may therefore approximate $\delta Y^{(\mathrm{surf})}_{\mathbf{k}_\perp}(s)$ by its instantaneous value $\delta Y^{(\mathrm{surf})}_{\mathbf{k}_\perp}(t)$, giving 
\begin{equation} 
\begin{split}
    I_1(t) &\approx 
    \frac{\bar{r}}{\bar{Y}^{(\mathrm{surf})}} 
    \left( \int_{-\infty}^t e^{-\frac{\bar{r}(t-s)}{\bar{\theta}(1-\bar{\theta})}} ds \right) 
    \delta Y^{(\mathrm{surf})}_{\mathbf{k}_\perp}(t) \\
    &= \frac{\bar{\theta}(1-\bar{\theta})}{\bar{Y}^{(\mathrm{surf})}} \delta Y^{(\mathrm{surf})}_{\mathbf{k}_\perp}(t). 
\end{split}
\end{equation} 
Therefore, in the small-$k_\perp$ regime, $I_1(t)$ represents the quasi-static response of the surface coverage to the slowly varying mass-fraction fluctuation, with proportionality coefficient $\bar{\theta}(1-\bar{\theta})/\bar{Y}^{(\mathrm{surf})}$. 
Since the static structure factor is quadratic in the fluctuation amplitude, this quasi-static relation provides a theoretical motivation for the factor $\left[\bar{\theta}(1-\bar{\theta})/\bar{Y}^{(\mathrm{surf})}\right]^2$ underlying the scaling relation in Eq.~\eqref{eq:SFscaling}.

We emphasize, however, that this argument does not constitute a complete analytical derivation of the structure factor. 
Although the expressions for $I_1(t)$ and $I_2(t)$ are explicit, the history of $\delta Y_{\mathbf{k}_\perp}^{(\mathrm{surf})}$ entering $I_1(t)$ is itself governed by the coupled stochastic dynamics of the gas--surface system and can therefore be correlated with the noise entering $I_2(t)$.
Consequently, the full structure factor of $\delta\theta_{\mathbf{k}_\perp}$ cannot, in general, be obtained by treating $I_1$ and $I_2$ as statistically independent contributions.
Nevertheless, the small-$k_\perp$ time-scale separation provides a theoretical basis for the proportionality factor underlying the scaling relation in Eq.~\eqref{eq:SFscaling}.

\section{\label{sec:appendixNonreactive}Approximate Expression for $S_Y^{(\mathrm{surf})}$ in the Hypothetical Nonreactive System}

The main assumption of the hypothetical nonreactive system considered in this work is that the same concentration gradient $\overline{\nabla Y}$ is sustained in the absence of surface reactions.
To this end, for $Y$, we impose an inhomogeneous Neumann boundary condition at the surface (i.e., $\partial Y/\partial z = \lvert \overline{\nabla Y}\rvert$ at $z=0$), whereas we retain the Dirichlet boundary condition at the reservoir (i.e., $Y=Y^{(\mathrm{res})}$ at $z=L$).
For the velocity field, we assume no-slip boundary conditions.
Following the analytical approach of Ref.~\cite{ZarateRedondo2001}, we obtain an approximate expression for $S_Y^{(\mathrm{surf})}$.
To directly compare continuum and numerical structure factors, we evaluate the structure factor averaged across the bottom layer (i.e., $\frac{1}{\Delta z}\int_0^{\Delta z} dz \int_0^{\Delta z} dz'\; S_Y(k_\perp,z,z'))$.
The resulting expression reads (see Sec.~S2 in the Supplemental Material~\cite{SM} for details):
\begin{equation}
\label{eq:nonreactiveSYsurf}
\begin{aligned}
    &S_{Y}^{(\mathrm{surf})}(k_{\perp}) = S_{Y,\mathrm{eq}}^{(\mathrm{surf})} \\
    &+ \frac{k_BT \lvert\overline{\nabla Y} \rvert ^2}{\rho D}\frac{40320 k_{\perp}^2 L^7}{\pi^2 \Delta z}\sum_{N=1}^{\infty}\sum_{M=1}^{\infty}\frac{1}{((2N-1)^2+(2M-1)^2)\pi^2+8 k_{\perp} ^2L^2} \frac{W_N}{2N-1}\frac{W_M}{2M-1} \\
    & \times \left[\frac{1}{g(2N-1,k_{\perp} L)}+\frac{1}{g(2M-1,k_{\perp} L)}\right] \sin{\left(\frac{(2N-1)\pi\Delta z}{2L}\right)} \sin{\left(\frac{(2M-1)\pi\Delta z}{2L}\right)} \\
    &+ \frac{k_BT \lvert \overline{\nabla Y} \rvert^2}{\rho D}\frac{1774080 k_{\perp}^2L^7}{\pi^2\Delta z}\sum_{N=1}^{\infty}\sum_{M=1}^{\infty}\frac{1}{((2N-1)^2+(2M-1)^2)\pi^2+8 k_{\perp}^2L^2} \frac{\mathcal{W}_N}{2N-1}\frac{\mathcal{W}_M}{2M-1} \\
    &\times \left[\frac{1}{\mathcal{G}(2N-1,k_{\perp} L)}+\frac{1}{\mathcal{G}(2M-1,k_{\perp} L)}\right] \sin{\left(\frac{(2N-1)\pi\Delta z}{2L}\right)} \sin{\left(\frac{(2M-1)\pi\Delta z}{2L}\right)},
\end{aligned}
\end{equation}
where $\Delta z$ is the height of the bottom layer (equal to $\Delta x$ in our simulations), $\rho$ is the mass density of the gas (equal to $\bar{\rho}_\mathrm{tot}$), $\nu$ is the kinematic viscosity,
\begin{equation}
\begin{aligned}
    W_N &= 32\frac{(-1)^N[(2N-1)^2\pi^2-48]-12\pi(2N-1)}{(2N-1)^5\pi^5}, \\
    g(n,\kappa) &= D(\kappa^2+12)(4\kappa^2+n^2\pi^2)+4\nu(\kappa^4+24\kappa^2+504), \\
    \mathcal{W}_N &= -16\frac{(2N-1)\pi \left[24(2N-1)\pi + (-1)^N\left((2N-1)^2\pi^2-240\right)\right]-960}{(2N-1)^6\pi^6}, \\
    \mathcal{G}(n,\kappa) &= D(\kappa^2+44)(4\kappa^2+n^2\pi^2)+4\nu(\kappa^4+88\kappa^2+3960).
\end{aligned}
\end{equation}


\input{paper.bbl}



\section*{Supplementary material}

## S1. SUPPLEMENTARY STRUCTURE FACTOR SPECTRA

### S1.1. Mass Fraction Structure Factor Spectra in a Confined Fluid Adjacent to a Reactive Surface

To provide a gas-phase reference for the surface coverage structure factors discussed in the main text, Fig. S1 shows the mass fraction structure factor spectra in the confined fluid. The spectra are computed both over the bulk region ($S_Y^{(\mathrm{bulk})}$) and for the cross section immediately above the reactive surface ($S_Y^{(\mathrm{surf})}$).

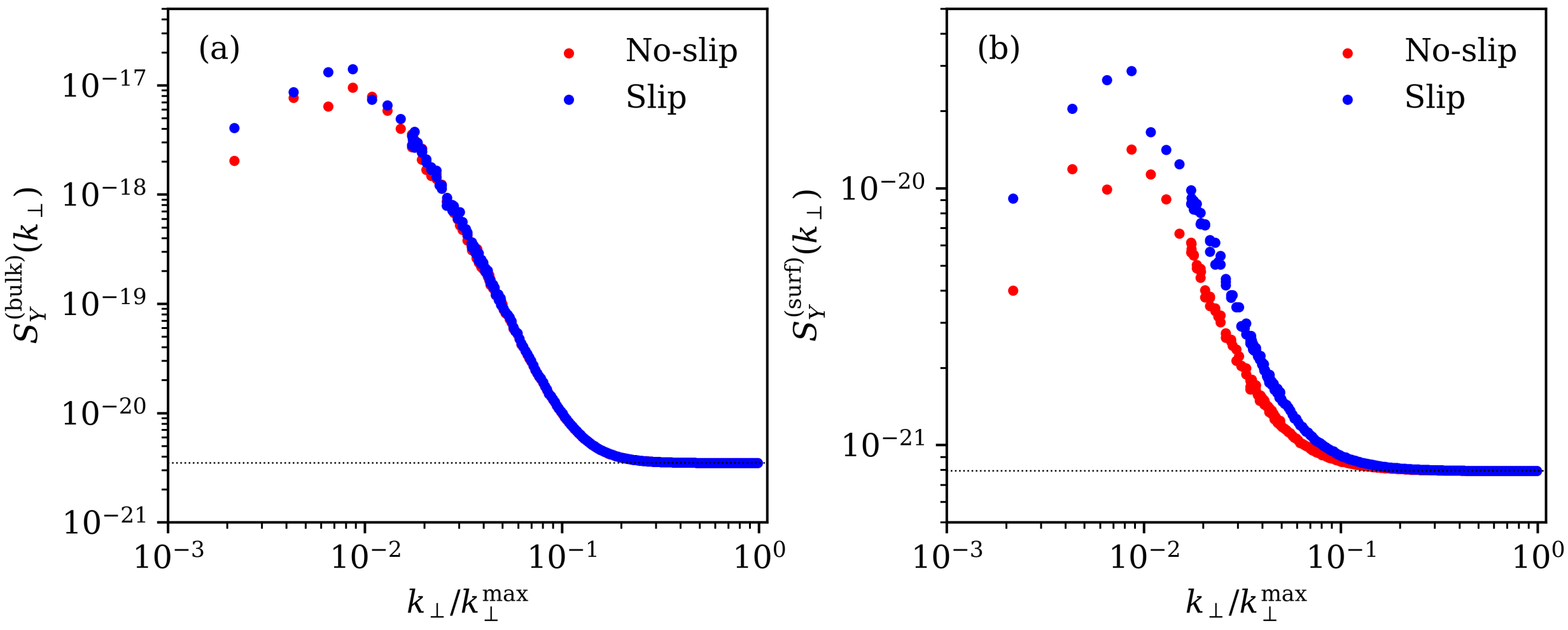


FIG. S1. Structure factor spectra of the mass fraction (a) over the bulk region, $S_Y^{(\mathrm{bulk})}(k_\perp)$, and (b) immediately above the surface, $S_Y^{(\mathrm{surf})}(k_\perp)$, obtained from three-dimensional simulations with slip and no-slip velocity boundary conditions at the reactive surface. The lateral wave number $k_\perp$ is normalized by its maximum value, $k_\perp^{\max} = 2\sqrt{2}/\Delta x$. The dotted lines denote the corresponding equilibrium structure factors and bin averaging is applied in the large-wave-number regime ($k_\perp/k_\perp^{\max} \geq 0.05$) to reduce the number of data points for clarity.

### S1.2. Comparison between Three-Dimensional and Quasi-Two-Dimensional Simulations

To confirm whether the quasi-two-dimensional geometry reproduces the surface coverage fluctuations observed in the three-dimensional system, Fig. S2 compares the corresponding surface coverage structure factor spectra. The comparison is performed for both slip and no-slip velocity boundary conditions at the reactive surface.

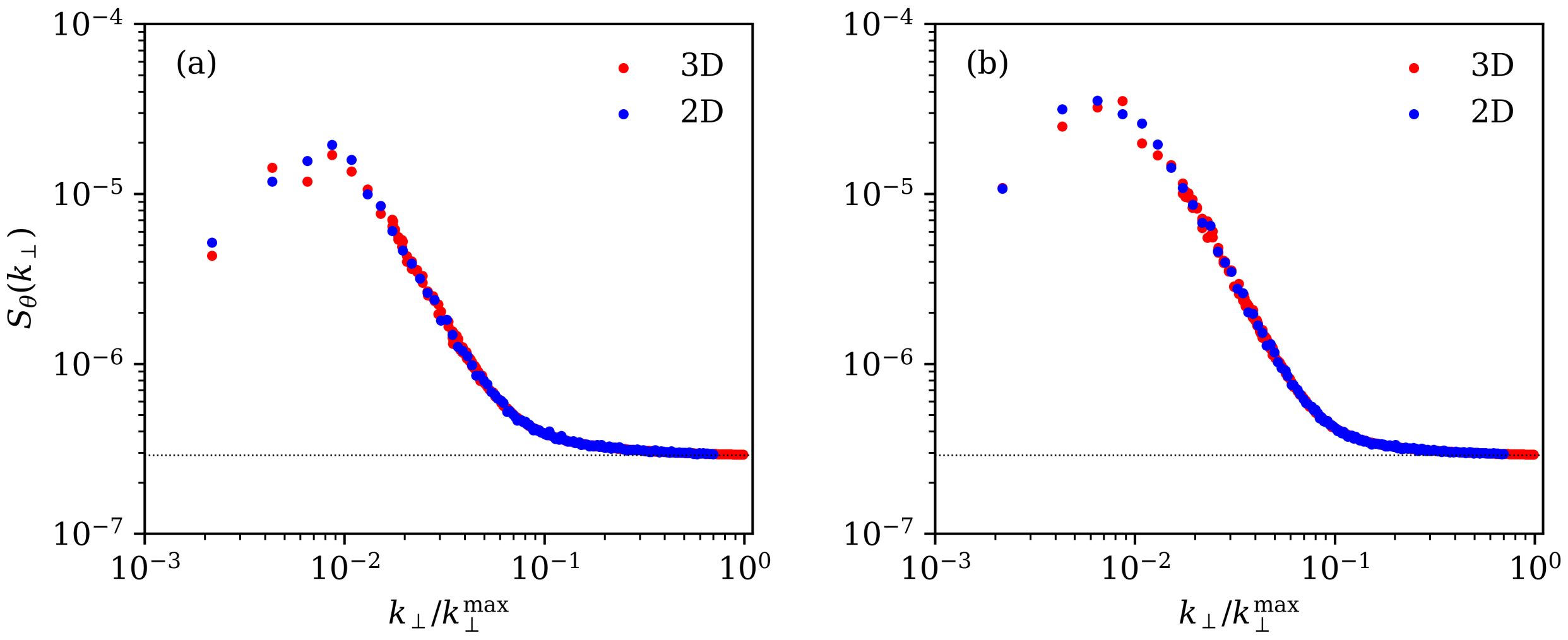


FIG. S2. Structure factor spectra of the surface coverage $S_\theta(k_\perp)$ obtained from simulations of three- and quasi-two-dimensional systems with (a) slip and (b) no-slip boundary conditions for velocity at the reactive surface. For both three- and quasi-two-dimensional simulation results, the lateral wave number $k_\perp$ is normalized by the three-dimensional maximum value, $k_\perp^{\max} = 2\sqrt{2}/\Delta x$, to enable a direct comparison between the two geometries. The dotted lines denote the corresponding equilibrium structure factors and bin averaging is applied in the large-wave-number region ($k_\perp/k_\perp^{\max} \geq 0.05$) to reduce the number of data points for clarity.

### S1.3. Surface Structure Factor Spectra with a Slip Velocity Boundary Condition

To assess the robustness of the results shown in Fig. 4 of the main text with respect to the choice of velocity boundary condition, Fig. S3 shows the surface structure factor obtained using a slip boundary condition at the reactive surface.

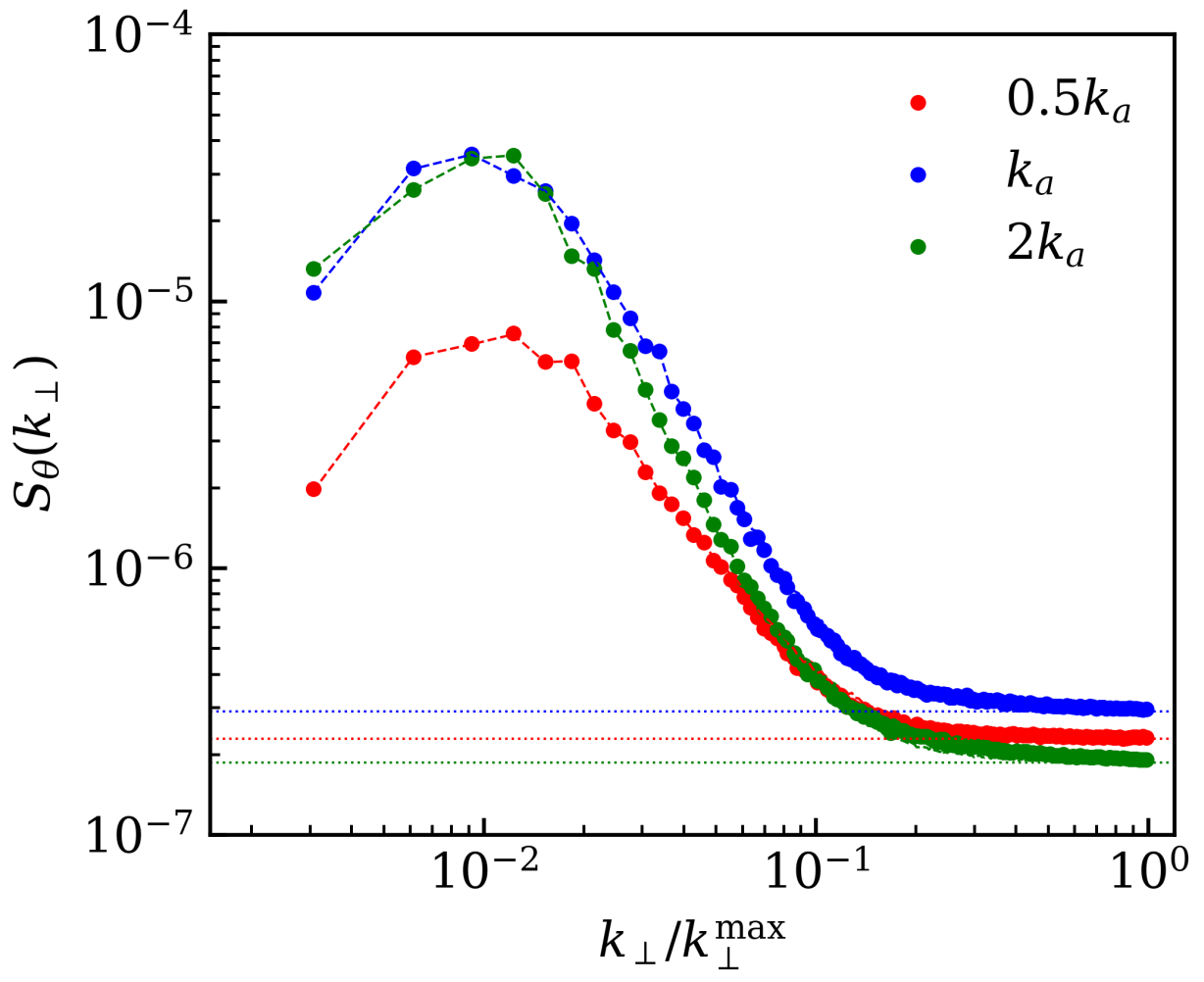


FIG. S3. Surface coverage structure factor spectra for different adsorption rate constants under the same imposed concentration gradient. The results labeled $k_a$ (blue) are obtained from a quasi-two-dimensional simulation with the parameter values listed in Table I of the main text. For the cases labeled $0.5k_a$ and $2k_a$ (red and green, respectively), the adsorption rate constant is varied while all other parameters, except $\bar{\theta}$, are held fixed. The corresponding values of $\bar{\theta}$ are adjusted to satisfy Eq. (2) in the main text. The dashed lines show $S_\theta$ computed from Eq. (3) in the main text using the corresponding numerical results for $S_Y^{(\mathrm{surf})}$. The lateral wave number $k_\perp$ is normalized by its maximum value, $k_\perp^{\max} = 2/\Delta x$. The dotted lines denote the corresponding equilibrium structure factors, and bin averaging is applied in the large-wave-number regime ($k_\perp/k_\perp^{\max} \geq 0.05$) to reduce the number of data points for clarity. A slip boundary condition is imposed at the surface. The corresponding results obtained with a no-slip boundary condition are shown in Fig. 4 of the main text.

### S1.4. Height Dependence of Mass Fraction Structure Factor Spectra in a Confined Fluid Adjacent to a Reactive Surface

To clarify the origin of the breakdown of the weak-coupling approximation observed in Fig. 6 of the main text, we analyze the mass fraction structure factor $S_Y^{(z_\ell)}(k_\perp)$ at different heights above the reactive surface. Specifically, we compare $S_Y^{(z_\ell)}$ obtained from the reactive system with the corresponding predictions for a hypothetical nonreactive system sustaining the same imposed concentration gradient. As shown in Fig. S4, agreement improves with increasing height, consistent with a reduced influence of surface reactions away from the interface.

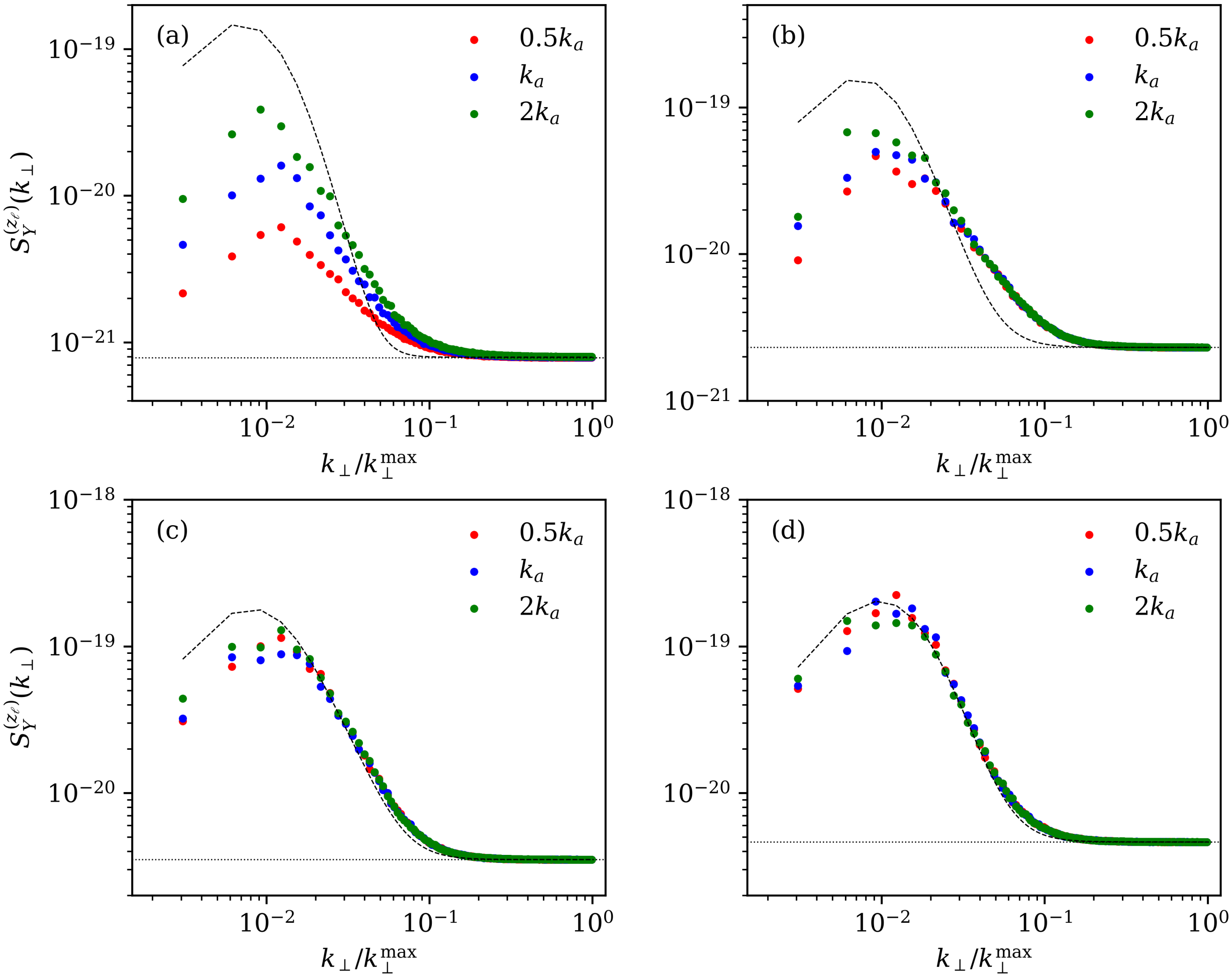


FIG. S4. Mass fraction structure factor spectra $S_Y^{(z_\ell)}(k_\perp)$ evaluated at different heights $z_\ell$ above the reactive surface: (a) $z_\ell \simeq 0$ (immediately above the surface), (b) $z_\ell \simeq L/8$, (c) $z_\ell \simeq L/4$, and (d) $z_\ell \simeq L/2$. Colored circles show simulation results from the reactive system, while dashed lines indicate predictions for the hypothetical nonreactive system under the same imposed concentration gradient. The dotted lines denote the corresponding equilibrium structure factors. The results labeled $k_a$ (blue) are obtained from a quasi-two-dimensional simulation with the parameter values listed in Table I of the main text. Parameter variations for the cases labeled $0.5k_a$ and $2k_a$ (red and green, respectively) are the same as those described in Fig. 4 of the main text. The lateral wave number $k_\perp$ is normalized by its maximum value, $k_\perp^{\max} = 2/\Delta x$, and bin averaging is applied in the large-wave-number regime ($k_\perp/k_\perp^{\max} \geq 0.05$) to reduce the number of data points for clarity. A no-slip velocity boundary condition is imposed at the surface.

## S2. DERIVATION OF AN APPROXIMATE STRUCTURE FACTOR FOR A HYPOTHETICAL NONREACTIVE SYSTEM

In this section, we derive an approximate analytic expression for $S_Y^{(\mathrm{surf})}$, Eq. (6) in the main text, for a hypothetical nonreactive system of a confined fluid sustaining a linear concentration gradient in the $z$ direction. To obtain this expression, we first derive the more general dynamic structure factor $S_Y(\omega, k_\perp, z, z')$, defined as

$$\langle \delta Y(\omega, k_\perp, z) \delta Y^*(\omega', k'_\perp, z') \rangle = S_Y(\omega, k_\perp, z, z')(2\pi)^3 \delta(\omega - \omega')\delta(k_\perp - k'_\perp). \tag{S1}$$

Here, the mass fraction fluctuation $\delta Y(\omega, k_\perp, z)$ denotes the deviation of the mass fraction $Y$ from its mean value $\overline{Y}(z)$. The variables $\omega$ and $k_\perp = (k_x, k_y)$ are Fourier variables corresponding to time $t$ and the lateral coordinates $(x, y)$, respectively.

Our derivation closely follows the analytic approach developed in Ref. [1]. However, because we consider different physical variables (mass fraction rather than temperature) and different boundary conditions (Neumann rather than Dirichlet), the resulting algebra and expressions differ. In Sec. S2.1, we describe the setup and assumptions of the hypothetical nonreactive system and outline the key differences from the system considered in Ref. [1]. In Sec. S2.2, we obtain a series representation of $\delta Y(\omega, k_\perp, z)$. In Sec. S2.3, we derive series expressions for various structure factors, including $S_Y(\omega, k_\perp, z, z')$ and $S_Y^{(\mathrm{surf})}(k_\perp)$.

### S2.1. System Setup and Governing Equations

We consider a system confined between two parallel planes at $z = 0$ and $z = L$. A constant concentration gradient $\overline{\nabla Y}$ in the mass fraction $Y$ is imposed in the $z$ direction. This gradient is imposed by an inhomogeneous Neumann condition $\partial Y/\partial z = |\overline{\nabla Y}|$ at $z = 0$ and an inhomogeneous Dirichlet condition $Y = Y_L$ at $z = L$. For the mean mass fraction $\bar{Y}(z) = Y_L - |\overline{\nabla Y}|(L - z)$, we define the mass fraction fluctuation $\delta Y(x, y, z) = Y(x, y, z) - \bar{Y}(z)$. For the velocity field, we assume no-slip boundary conditions.

The velocity component $v_\parallel$ parallel to the concentration gradient and the mass fraction fluctuation $\delta Y$ satisfy the following linearized equations:

$$\begin{aligned} \partial_t(\nabla^2 v_\parallel) &= \nu \nabla^2(\nabla^2 v_\parallel) + \rho^{-1}\{\nabla \times \nabla \times (\nabla \cdot \widetilde{\Pi})\}_\parallel, \\ \partial_t(\delta Y) &= -|\overline{\nabla Y}|\, v_\parallel + D\nabla^2(\delta Y). \end{aligned} \tag{S2}$$

Here, $\nu$ is the kinematic viscosity; $\rho$ is the mass density; $D$ is the binary diffusion coefficient; $\widetilde{\Pi}$ is the stochastic viscous flux tensor [1]; and the subscript $\|$ denotes the component parallel to the concentration gradient. As in Ref. [1], we neglect the stochastic flux in the equation for $\delta Y$ and retain the stochastic flux only in the equation for $v_\|$. This simplifies the algebra and enables us to focus on the nonequilibrium enhancement due to mode coupling. Accordingly, the structure factors derived in Sec. S2.3 contain only the nonequilibrium contribution; the full structure factors can be obtained by adding the corresponding equilibrium contributions.

The Fourier transform of Eq. (S2) with respect to $t$ and $(x, y)$ yields

$$\begin{bmatrix} i\omega\left(\frac{d^2}{dz^2}-k_\perp^2\right)-\nu\left(\frac{d^2}{dz^2}-k_\perp^2\right)^2 & 0 \\ |\overline{\nabla Y}| & i\omega - D\left(\frac{d^2}{dz^2}-k_\perp^2\right)\end{bmatrix}\begin{bmatrix} v_\|(\omega,k_\perp,z) \\ \delta Y(\omega,k_\perp,z)\end{bmatrix} = \begin{bmatrix} F_1(\omega,k_\perp,z) \\ 0\end{bmatrix}. \tag{S3}$$

Here, $F_1(\omega, k_\perp, z)$ is given in terms of the Fourier components of the stochastic viscous flux tensor, as derived in Ref. [1]:

$$\begin{aligned} F_1(\omega,k_\perp,z) =& \frac{i}{\rho}\left[\frac{d^2}{dz^2}\left(k_x\hat{\Pi}_{zx}+k_y\hat{\Pi}_{zy}\right)+k_\perp^2\left(k_x\hat{\Pi}_{xz}+k_y\hat{\Pi}_{yz}\right)\right] \\ &+\frac{1}{\rho}\frac{d}{dz}\left[k_\perp^2\hat{\Pi}_{zz}-\left\{k_x^2\hat{\Pi}_{xx}+k_y^2\hat{\Pi}_{yy}+k_xk_y(\hat{\Pi}_{xy}+\hat{\Pi}_{yx})\right\}\right]\end{aligned} \tag{S4}$$

with

$$\langle\hat{\Pi}_{ij}(\omega,k_\perp,z)\hat{\Pi}^*_{kl}(\omega',k'_\perp,z')\rangle = 2k_BT\eta(\delta_{ik}\delta_{jl}+\delta_{il}\delta_{jk})(2\pi)^3\times\delta(\omega-\omega')\delta(k_\perp-k'_\perp)\delta(z-z'), \tag{S5}$$

where $\eta = \rho\nu$ is the shear viscosity. The boundary conditions for $\delta Y$ and $v_\|$ are

$$\frac{d}{dz}\delta Y(\omega,k_\perp,z) = 0 \quad \text{at } z = 0, \tag{S6a}$$

$$\delta Y(\omega,k_\perp,z) = 0 \quad \text{at } z = L, \tag{S6b}$$

$$v_\|(\omega,k_\perp,z) = 0 \quad \text{at } z = 0,\ L, \tag{S6c}$$

$$\frac{d}{dz}v_\|(\omega,k_\perp,z) = 0 \quad \text{at } z = 0,\ L. \tag{S6d}$$

Eqs. (S3) and (S6) constitute the governing equations of the system. In Sec. S2.2, we derive series representations of $v_\|$ and $\delta Y$. Before proceeding, we summarize the main differences between this system and that in Ref. [1]: (i) Ref. [1] considers temperature fluctuations, whereas we consider mass fraction (or concentration) fluctuations; (ii) the velocity boundary conditions are identical, but for the scalar field Ref. [1] imposes a Dirichlet condition at $z = 0$, whereas we impose a Neumann condition at $z = 0$.

### S2.2. Series Expansion and Solution

We derive series representations of $v_\parallel(\omega, k_\perp, z)$ and $\delta Y(\omega, k_\perp, z)$ that satisfy Eqs. (S3) and (S6). Since Eq. (S3) is derived under the Stokes limit, $\delta Y$ depends on $v_\parallel$, but not vice versa; we therefore first solve for $v_\parallel$ and then for $\delta Y$.

#### 1. $v_\parallel(\omega, k_\perp, z)$

Since we impose no-slip boundary conditions, we approximate the $z$-component of the velocity field using a zeroth-order Galerkin polynomial [1] that satisfies the boundary conditions (S6c)–(S6d):

$$v_\parallel(\omega, k_\perp, z) = v_{\parallel,0}(\omega, k_\perp) \left( \frac{z}{L} - \frac{z^2}{L^2} \right)^2 . \tag{S7}$$

Substituting Eq. (S7) into Eq. (S3) and projecting onto the Galerkin mode yields

$$v_{\parallel,0}(\omega, k_\perp) \int_0^L \left( \frac{z}{L} - \frac{z^2}{L^2} \right)^2 \left[ i\omega \left( \frac{d^2}{dz^2} - k_\perp^2 \right) - \nu \left( \frac{d^2}{dz^2} - k_\perp^2 \right)^2 \right] \left( \frac{z}{L} - \frac{z^2}{L^2} \right)^2 dz = G(\omega, k_\perp), \tag{S8}$$

where

$$G(\omega, k_\perp) = \int_0^L \left( \frac{z}{L} - \frac{z^2}{L^2} \right)^2 F_1(\omega, k_\perp, z)\, dz. \tag{S9}$$

Hence, the amplitude $v_{\parallel,0}(\omega, k_\perp)$ is given by

$$v_{\parallel,0}(\omega, k_\perp) = -\frac{630 L^3 G(\omega, k_\perp)}{i\omega L^2 (k_\perp^2 L^2 + 12) + \nu (k_\perp^4 L^4 + 24 k_\perp^2 L^2 + 504)}. \tag{S10}$$

#### 2. $\delta Y(\omega, k_\perp, z)$

We now expand $\delta Y$ in a Fourier cosine basis on the interval $[0, L]$, which is compatible with the boundary conditions (S6a)–(S6b):

$$\delta Y(\omega, k_\perp, z) = \sum_{N=1}^{\infty} B_N(\omega, k_\perp) \cos\left( \frac{(2N-1)\pi z}{2L} \right). \tag{S11}$$

To maintain a consistent representation, we also project $v_\parallel$ onto the same basis and express $v_\parallel$ and $\delta Y$ as

$$\begin{bmatrix} v_\parallel(\omega, k_\perp, z) \\ \delta Y(\omega, k_\perp, z) \end{bmatrix} = \sum_{N=1}^{\infty} \begin{bmatrix} v_{\parallel,0}(\omega, k_\perp) W_N \\ B_N(\omega, k_\perp) \end{bmatrix} \cos\left( \frac{(2N-1)\pi z}{2L} \right), \tag{S12}$$

where the coefficients $W_N$ are obtained by expanding the Galerkin polynomial $(z/L - z^2/L^2)^2$ in the Fourier cosine basis:

$$W_N = 32\frac{(-1)^N[(2N-1)^2\pi^2 - 48] - 12\pi(2N-1)}{(2N-1)^5\pi^5}. \tag{S13}$$

By substituting Eq. (S12) into Eq. (S3), the coefficients $B_N(\omega, k_\perp)$ are found to be

$$B_N(\omega, k_\perp) = -\frac{|\overline{\nabla Y}|v_{\|,0}(\omega, k_\perp)W_N}{i\omega + D\left[\frac{(2N-1)^2\pi^2}{4L^2} + k_\perp^2\right]}. \tag{S14}$$

Hence, we obtain the expansion of $\delta Y$:

$$\begin{aligned}\delta Y(\omega, k_\perp, z) = &-\frac{630L^3\, G(\omega, k_\perp)}{i\omega L^2(k_\perp^2 L^2 + 12) + \nu(k_\perp^4 L^4 + 24k_\perp^2 L^2 + 504)} \\ &\times \sum_{N=1}^{\infty} \frac{-W_N|\overline{\nabla Y}|}{i\omega + D\left(\frac{(2N-1)^2\pi^2}{4L^2} + k_\perp^2\right)} \cos\left(\frac{(2N-1)\pi z}{2L}\right).\end{aligned} \tag{S15}$$

Before proceeding to Sec. S2.3 to obtain the structure factors of $\delta Y$, we evaluate the correlation function of $G(\omega, k_\perp)$, which is required to compute the structure factors. Using Eq. (S4) and Eq. (S5), the correlation $\langle G(\omega, k_\perp)G^*(\omega', k'_\perp)\rangle$ is given by

$$\begin{aligned}&\langle G(\omega, k_\perp)G^*(\omega', k'_\perp)\rangle \\ &= \frac{2\nu k_B T}{\rho}k_\perp^2(2\pi)^3\delta(\omega-\omega')\delta(k_\perp - k'_\perp) \times \int_0^L\int_0^L \left(\frac{z}{L} - \frac{z^2}{L^2}\right)^2\left(\frac{z'}{L} - \frac{z'^2}{L^2}\right)^2 \\ &\quad\times\left[k_\perp^4 + k_\perp^2\left(\frac{d^2}{dz^2} + \frac{d^2}{dz'^2} + 4\frac{d}{dz}\frac{d}{dz'}\right) + \frac{d^2}{dz^2}\frac{d^2}{dz'^2}\right]\delta(z-z')\,dz\,dz',\end{aligned} \tag{S16}$$

which yields

$$\langle G(\omega, k_\perp)G^*(\omega', k'_\perp)\rangle = \frac{2\nu k_B T}{\rho}k_\perp^2\frac{k_\perp^4 L^4 + 24k_\perp^2 L^2 + 504}{630L^3}(2\pi)^3 \times \delta(\omega-\omega')\delta(k_\perp - k'_\perp). \tag{S17}$$

### S2.3. Structure Factors

Using the series representations obtained above, we now derive series expressions for various structure factors of $\delta Y$. We first derive the dynamic and static structure factors, $S_Y(\omega, k_\perp, z, z')$ and $S_Y(k_\perp, z, z')$ for $z, z' \in [0, L]$. We then obtain the bulk structure factor, $S_Y^{(\text{bulk})}(k_\perp)$, and the cross-sectional structure factor, $S_Y^{(z_\ell)}(k_\perp)$. Finally, we obtain the structure factor immediately above the plane $z = 0$, $S_Y^{(\text{surf})}(k_\perp)$, whose expression is given in Eq. (5) of the End Matter.

### 1. *Dynamic Structure Factor* $S_Y(\omega, k_\perp, z, z')$ *and Static Structure Factor* $S_Y(k_\perp, z, z')$

Substituting Eq. (S15) into Eq. (S1) and using Eq. (S17), the dynamic structure factor $S_Y(\omega, k_\perp, z, z')$ is given by

$$S_Y(\omega, k_\perp, z, z') = \frac{(630L^3)^2 \langle G(\omega, k_\perp) G^*(\omega, k_\perp)\rangle/(2\pi)^3}{[\omega L^2(k_\perp^2 L^2 + 12)]^2 + [\nu(k_\perp^4 L^4 + 24k_\perp^2 L^2 + 504)]^2} |\overline{\nabla Y}|^2 \times \sum_{N=1}^{\infty}\sum_{M=1}^{\infty} \frac{W_N \cos\left(\frac{(2N-1)\pi z}{2L}\right)}{i\omega + D\left(\frac{(2N-1)^2\pi^2}{4L^2} + k_\perp^2\right)} \frac{W_M \cos\left(\frac{(2M-1)\pi z'}{2L}\right)}{-i\omega + D\left(\frac{(2M-1)^2\pi^2}{4L^2} + k_\perp^2\right)}. \quad \text{(S18)}$$

Integrating over the frequency $\omega$, the static structure factor $S_Y(k_\perp, z, z')$ is obtained by

$$S_Y(k_\perp, z, z') = \frac{k_B T |\overline{\nabla Y}|^2}{\rho D} 10080 k_\perp^2 L^3 \sum_{N=1}^{\infty}\sum_{M=1}^{\infty} \frac{W_N W_M \cos\left(\frac{(2N-1)\pi z}{2L}\right)\cos\left(\frac{(2M-1)\pi z'}{2L}\right)}{((2N-1)^2 + (2M-1)^2)\pi^2 + 8k_\perp^2 L^2} \times \left[\frac{1}{g(2N-1, k_\perp L)} + \frac{1}{g(2M-1, k_\perp L)}\right], \quad \text{(S19)}$$

where

$$g(n, \kappa) = D(\kappa^2 + 12)(4\kappa^2 + n^2\pi^2) + 4\nu(\kappa^4 + 24\kappa^2 + 504). \quad \text{(S20)}$$

### 2. *Bulk Structure Factor* $S_Y^{(\mathrm{bulk})}(k_\perp)$

From the static structure factor $S_Y(k_\perp, z, z')$, the scattering intensity $S_Y^{(\mathrm{bulk})}(k_\perp, k_\parallel)$ can be computed as [1]

$$S_Y^{(\mathrm{bulk})}(k_\perp, k_\parallel) = \frac{1}{L}\int_0^L \int_0^L e^{-ik_\parallel(z-z')} S_Y(k_\perp, z, z') dz dz' \quad \text{(S21)}$$

and, in the limit $k_\parallel \to 0$, we obtain

$$S_Y^{(\mathrm{bulk})}(k_\perp) = \frac{k_B T |\overline{\nabla Y}|^2}{\rho D} \frac{40320 k_\perp^2 L^6}{\pi^2} \sum_{N=1}^{\infty}\sum_{M=1}^{\infty} \frac{1}{((2N-1)^2 + (2M-1)^2)\pi^2 + 8k_\perp^2 L^2} \times (-1)^{N+M} \frac{W_N}{2N-1}\frac{W_M}{2M-1}\left[\frac{1}{g(2N-1, k_\perp L)} + \frac{1}{g(2M-1, k_\perp L)}\right]. \quad \text{(S22)}$$

### 3. *Cross-sectional Structure Factor* $S_Y^{(z_\ell)}(k_\perp)$

While the bulk structure factor corresponds to an average over the entire system, the cross-sectional structure factor requires resolving the intensity within individual slices

of thickness $\Delta z$. From the static structure factor $S_Y(k_\perp, z, z')$, the scattering intensity $S_Y^{(z_\ell)}(k_\perp, k_\parallel)$ can be computed as

$$S_Y^{(z_\ell)}(k_\perp, k_\parallel) = \frac{1}{\Delta z}\int_{(\ell-1)\Delta z}^{\ell\Delta z}\int_{(\ell-1)\Delta z}^{\ell\Delta z} e^{-ik_\parallel(z-z')} S_Y(k_\perp, z, z')\, dz\, dz', \tag{S23}$$

where $\ell$ denotes the discrete slice index, with $\ell = 1$ corresponding to the bottom slice adjacent to the plane $z = 0$. In the limit $k_\parallel \to 0$, we obtain

$$\begin{aligned} S_Y^{(z_\ell)}(k_\perp) =& \frac{k_B T|\overline{\nabla Y}|^2}{\rho D}\frac{40320 k_\perp^2 L^7}{\pi^2 \Delta z}\sum_{N=1}^{\infty}\sum_{M=1}^{\infty}\frac{1}{((2N-1)^2+(2M-1)^2)\pi^2+8k_\perp^2 L^2} \\ &\times \frac{W_N}{2N-1}\frac{W_M}{2M-1}\left[\frac{1}{g(2N-1,k_\perp L)}+\frac{1}{g(2M-1,k_\perp L)}\right] \\ &\times\left[\sin\left(\frac{(2N-1)\pi}{2L}\ell\Delta z\right)-\sin\left(\frac{(2N-1)\pi}{2L}(\ell-1)\Delta z\right)\right] \\ &\times\left[\sin\left(\frac{(2M-1)\pi}{2L}\ell\Delta z\right)-\sin\left(\frac{(2M-1)\pi}{2L}(\ell-1)\Delta z\right)\right]. \end{aligned} \tag{S24}$$

To compute $S_Y^{(z_\ell)}$ more accurately, one can consider the first-order Galerkin approximation for $v_\parallel$ with the following basis [1]:

$$\phi_0(z) = \left(\frac{z}{L}-\frac{z^2}{L^2}\right)^2, \quad \phi_1(z) = \left(\frac{z}{L}-\frac{z^2}{L^2}\right)^2\left(\frac{1}{2}-\frac{z}{L}\right). \tag{S25}$$

The resulting first-order correction is given by

$$\begin{aligned} \Delta S_Y^{(z_\ell)}(k_\perp) =& \frac{k_B T|\overline{\nabla Y}|^2}{\rho D}\frac{1774080 k_\perp^2 L^7}{\pi^2 \Delta z}\sum_{N=1}^{\infty}\sum_{M=1}^{\infty}\frac{1}{((2N-1)^2+(2M-1)^2)\pi^2+8k_\perp^2 L^2} \\ &\times \frac{\mathcal{W}_N}{2N-1}\frac{\mathcal{W}_M}{2M-1}\left[\frac{1}{\mathcal{G}(2N-1,k_\perp L)}+\frac{1}{\mathcal{G}(2M-1,k_\perp L)}\right] \\ &\times\left[\sin\left(\frac{(2N-1)\pi}{2L}\ell\Delta z\right)-\sin\left(\frac{(2N-1)\pi}{2L}(\ell-1)\Delta z\right)\right] \\ &\times\left[\sin\left(\frac{(2M-1)\pi}{2L}\ell\Delta z\right)-\sin\left(\frac{(2M-1)\pi}{2L}(\ell-1)\Delta z\right)\right], \end{aligned} \tag{S26}$$

where

$$\mathcal{W}_N = -16\frac{(2N-1)\pi\left[24(2N-1)\pi+(-1)^N\left((2N-1)^2\pi^2-240\right)\right]-960}{(2N-1)^6\pi^6}, \tag{S27}$$

$$\mathcal{G}(n,\kappa) = D(\kappa^2+44)(4\kappa^2+n^2\pi^2)+4\nu(\kappa^4+88\kappa^2+3960). \tag{S28}$$

As mentioned in Sec. S2.1, these results do not include the equilibrium contribution, which is given by

$$S_{Y,\mathrm{eq}}^{(z_\ell)} = \frac{m}{\rho}\bar{Y}^{(z_\ell)}(1-\bar{Y}^{(z_\ell)}), \tag{S29}$$

where $\bar{Y}^{(z_\ell)}$ denotes the mean mass fraction in the slice. Therefore, our final approximate expression for the cross-sectional structure factor is given by $S_{Y,\mathrm{eq}}^{(z_\ell)} + S_Y^{(z_\ell)} + \Delta S_Y^{(z_\ell)}$.

### 4. *Structure Factor Immediately above the Surface, $S_Y^{(\mathrm{surf})}(k_\perp)$*

The structure factor immediately above the plane $z = 0$, denoted by $S_Y^{(\mathrm{surf})}(k_\perp)$ in the main text, is obtained from the cross-sectional structure factor with $\ell = 1$. Hence,

$$S_Y^{(\mathrm{surf})} \approx S_{Y,\mathrm{eq}}^{(z_1)} + S_Y^{(z_1)} + \Delta S_Y^{(z_1)}. \tag{S30}$$


[1] J. M. Ortiz de Zárate and L. M. Redondo, Eur. Phys. J. B **21**, 135 (2001).



\end{document}

%% file: paper.bbl
%

%% file: paper.bbl
\begin{thebibliography}{18}%
\makeatletter
\providecommand \@ifxundefined [1]{%
 \@ifx{#1\undefined}
}%
\providecommand \@ifnum [1]{%
 \ifnum #1\expandafter \@firstoftwo
 \else \expandafter \@secondoftwo
 \fi
}%
\providecommand \@ifx [1]{%
 \ifx #1\expandafter \@firstoftwo
 \else \expandafter \@secondoftwo
 \fi
}%
\providecommand \natexlab [1]{#1}%
\providecommand \enquote  [1]{``#1''}%
\providecommand \bibnamefont  [1]{#1}%
\providecommand \bibfnamefont [1]{#1}%
\providecommand \citenamefont [1]{#1}%
\providecommand \href@noop [0]{\@secondoftwo}%
\providecommand \href [0]{\begingroup \@sanitize@url \@href}%
\providecommand \@href[1]{\@@startlink{#1}\@@href}%
\providecommand \@@href[1]{\endgroup#1\@@endlink}%
\providecommand \@sanitize@url [0]{\catcode `\\12\catcode `\$12\catcode `\&12\catcode `\#12\catcode `\^12\catcode `\_12\catcode `\%12\relax}%
\providecommand \@@startlink[1]{}%
\providecommand \@@endlink[0]{}%
\providecommand \url  [0]{\begingroup\@sanitize@url \@url }%
\providecommand \@url [1]{\endgroup\@href {#1}{\urlprefix }}%
\providecommand \urlprefix  [0]{URL }%
\providecommand \Eprint [0]{\href }%
\providecommand \doibase [0]{https://doi.org/}%
\providecommand \selectlanguage [0]{\@gobble}%
\providecommand \bibinfo  [0]{\@secondoftwo}%
\providecommand \bibfield  [0]{\@secondoftwo}%
\providecommand \translation [1]{[#1]}%
\providecommand \BibitemOpen [0]{}%
\providecommand \bibitemStop [0]{}%
\providecommand \bibitemNoStop [0]{.\EOS\space}%
\providecommand \EOS [0]{\spacefactor3000\relax}%
\providecommand \BibitemShut  [1]{\csname bibitem#1\endcsname}%
\let\auto@bib@innerbib\@empty
\bibitem [{\citenamefont {Vailati}\ and\ \citenamefont {Giglio}(1997)}]{VailatiGiglio1997}%
  \BibitemOpen
  \bibfield  {author} {\bibinfo {author} {\bibfnamefont {A.}~\bibnamefont {Vailati}}\ and\ \bibinfo {author} {\bibfnamefont {M.}~\bibnamefont {Giglio}},\ }\bibfield  {title} {\bibinfo {title} {Giant fluctuations in a free diffusion process},\ }\href {https://doi.org/10.1038/36803} {\bibfield  {journal} {\bibinfo  {journal} {Nature}\ }\textbf {\bibinfo {volume} {390}},\ \bibinfo {pages} {262} (\bibinfo {year} {1997})}\BibitemShut {NoStop}%
\bibitem [{\citenamefont {Brogioli}\ \emph {et~al.}(2000)\citenamefont {Brogioli}, \citenamefont {Vailati},\ and\ \citenamefont {Giglio}}]{BrogioliVailatiGiglio2000}%
  \BibitemOpen
  \bibfield  {author} {\bibinfo {author} {\bibfnamefont {D.}~\bibnamefont {Brogioli}}, \bibinfo {author} {\bibfnamefont {A.}~\bibnamefont {Vailati}},\ and\ \bibinfo {author} {\bibfnamefont {M.}~\bibnamefont {Giglio}},\ }\bibfield  {title} {\bibinfo {title} {Universal behavior of nonequilibrium fluctuations in free diffusion processes},\ }\href {https://doi.org/10.1103/PhysRevE.61.R1} {\bibfield  {journal} {\bibinfo  {journal} {Phys.\ Rev.\ E}\ }\textbf {\bibinfo {volume} {61}},\ \bibinfo {pages} {R1} (\bibinfo {year} {2000})}\BibitemShut {NoStop}%
\bibitem [{\citenamefont {Vailati}\ and\ \citenamefont {Giglio}(1998)}]{VailatiGiglio1998}%
  \BibitemOpen
  \bibfield  {author} {\bibinfo {author} {\bibfnamefont {A.}~\bibnamefont {Vailati}}\ and\ \bibinfo {author} {\bibfnamefont {M.}~\bibnamefont {Giglio}},\ }\bibfield  {title} {\bibinfo {title} {Nonequilibrium fluctuations in time-dependent diffusion processes},\ }\href {https://doi.org/10.1103/PhysRevE.58.4361} {\bibfield  {journal} {\bibinfo  {journal} {Phys.\ Rev.\ E}\ }\textbf {\bibinfo {volume} {58}},\ \bibinfo {pages} {4361} (\bibinfo {year} {1998})}\BibitemShut {NoStop}%
\bibitem [{\citenamefont {Ortiz~de Z{\'{a}}rate}\ and\ \citenamefont {Sengers}(2006)}]{ZarateSengers2006}%
  \BibitemOpen
  \bibfield  {author} {\bibinfo {author} {\bibfnamefont {J.~M.}\ \bibnamefont {Ortiz~de Z{\'{a}}rate}}\ and\ \bibinfo {author} {\bibfnamefont {J.~V.}\ \bibnamefont {Sengers}},\ }\href@noop {} {\emph {\bibinfo {title} {Hydrodynamic Fluctuations in Fluids and Fluid Mixtures}}}\ (\bibinfo  {publisher} {Elsevier Science},\ \bibinfo {year} {2006})\BibitemShut {NoStop}%
\bibitem [{\citenamefont {Landau}\ and\ \citenamefont {Lifschitz}(1987)}]{LandauLifschitz1987}%
  \BibitemOpen
  \bibfield  {author} {\bibinfo {author} {\bibfnamefont {L.~D.}\ \bibnamefont {Landau}}\ and\ \bibinfo {author} {\bibfnamefont {E.~M.}\ \bibnamefont {Lifschitz}},\ }\href@noop {} {\emph {\bibinfo {title} {Fluid Mechanics}}},\ \bibinfo {edition} {2nd}\ ed.\ (\bibinfo  {publisher} {Pergamon Press},\ \bibinfo {year} {1987})\BibitemShut {NoStop}%
\bibitem [{\citenamefont {Garcia}\ \emph {et~al.}()\citenamefont {Garcia}, \citenamefont {Bell}, \citenamefont {Nonaka}, \citenamefont {Srivastava}, \citenamefont {Ladiges},\ and\ \citenamefont {Kim}}]{GarciaBellNonakaSrivastavaLadigesKim2026}%
  \BibitemOpen
  \bibfield  {author} {\bibinfo {author} {\bibfnamefont {A.~L.}\ \bibnamefont {Garcia}}, \bibinfo {author} {\bibfnamefont {J.~B.}\ \bibnamefont {Bell}}, \bibinfo {author} {\bibfnamefont {A.}~\bibnamefont {Nonaka}}, \bibinfo {author} {\bibfnamefont {I.}~\bibnamefont {Srivastava}}, \bibinfo {author} {\bibfnamefont {D.}~\bibnamefont {Ladiges}},\ and\ \bibinfo {author} {\bibfnamefont {C.}~\bibnamefont {Kim}},\ }\bibfield  {title} {\bibinfo {title} {An introduction to computational fluctuating hydrodynamics},\ }\href@noop {} {\bibfield  {journal} {\bibinfo  {journal} {SIAM Rev.}\ }}\bibinfo {note} {To appear},\ \Eprint {https://arxiv.org/abs/arXiv:2406.12157} {arXiv:2406.12157} \BibitemShut {NoStop}%
\bibitem [{\citenamefont {Donev}\ \emph {et~al.}(2011)\citenamefont {Donev}, \citenamefont {Bell}, \citenamefont {de~la Fuente},\ and\ \citenamefont {Garcia}}]{DonevBellDeLaFuenteGarcia2011}%
  \BibitemOpen
  \bibfield  {author} {\bibinfo {author} {\bibfnamefont {A.}~\bibnamefont {Donev}}, \bibinfo {author} {\bibfnamefont {J.~B.}\ \bibnamefont {Bell}}, \bibinfo {author} {\bibfnamefont {A.}~\bibnamefont {de~la Fuente}},\ and\ \bibinfo {author} {\bibfnamefont {A.~L.}\ \bibnamefont {Garcia}},\ }\bibfield  {title} {\bibinfo {title} {Diffusive transport by thermal velocity fluctuations},\ }\href {https://doi.org/10.1103/PhysRevLett.106.204501} {\bibfield  {journal} {\bibinfo  {journal} {Phys.\ Rev.\ Lett.}\ }\textbf {\bibinfo {volume} {106}},\ \bibinfo {pages} {204501} (\bibinfo {year} {2011})}\BibitemShut {NoStop}%
\bibitem [{\citenamefont {Takacs}\ \emph {et~al.}(2011)\citenamefont {Takacs}, \citenamefont {Vailati}, \citenamefont {Cerbino}, \citenamefont {Mazzoni}, \citenamefont {Giglio},\ and\ \citenamefont {Cannell}}]{TakacsVailatiCerbinoMazzoniGiglioCannell2011}%
  \BibitemOpen
  \bibfield  {author} {\bibinfo {author} {\bibfnamefont {C.~J.}\ \bibnamefont {Takacs}}, \bibinfo {author} {\bibfnamefont {A.}~\bibnamefont {Vailati}}, \bibinfo {author} {\bibfnamefont {R.}~\bibnamefont {Cerbino}}, \bibinfo {author} {\bibfnamefont {S.}~\bibnamefont {Mazzoni}}, \bibinfo {author} {\bibfnamefont {M.}~\bibnamefont {Giglio}},\ and\ \bibinfo {author} {\bibfnamefont {D.~S.}\ \bibnamefont {Cannell}},\ }\bibfield  {title} {\bibinfo {title} {Thermal fluctuations in a layer of liquid cs$_2$ subjected to temperature gradients with and without the influence of gravity},\ }\href {https://doi.org/10.1103/PhysRevLett.106.244502} {\bibfield  {journal} {\bibinfo  {journal} {Phys.\ Rev.\ Lett.}\ }\textbf {\bibinfo {volume} {106}},\ \bibinfo {pages} {244502} (\bibinfo {year} {2011})}\BibitemShut {NoStop}%
\bibitem [{\citenamefont {Ortiz~de Z{\'{a}}rate}\ \emph {et~al.}(2015)\citenamefont {Ortiz~de Z{\'{a}}rate}, \citenamefont {Kirkpatrick},\ and\ \citenamefont {Sengers}}]{ZarateKirkpatrickSengers2015}%
  \BibitemOpen
  \bibfield  {author} {\bibinfo {author} {\bibfnamefont {J.~M.}\ \bibnamefont {Ortiz~de Z{\'{a}}rate}}, \bibinfo {author} {\bibfnamefont {T.~R.}\ \bibnamefont {Kirkpatrick}},\ and\ \bibinfo {author} {\bibfnamefont {J.~V.}\ \bibnamefont {Sengers}},\ }\bibfield  {title} {\bibinfo {title} {Non-equilibrium concentration fluctuations in binary liquids with realistic boundary conditions},\ }\href {https://doi.org/10.1140/epje/i2015-15099-x} {\bibfield  {journal} {\bibinfo  {journal} {Eur.\ Phys.\ J.\ E}\ }\textbf {\bibinfo {volume} {38}},\ \bibinfo {pages} {99} (\bibinfo {year} {2015})}\BibitemShut {NoStop}%
\bibitem [{\citenamefont {Croccolo}\ \emph {et~al.}(2016)\citenamefont {Croccolo}, \citenamefont {Ortiz~de Z{\'a}rate},\ and\ \citenamefont {Sengers}}]{CroccoloZarateSengers2016}%
  \BibitemOpen
  \bibfield  {author} {\bibinfo {author} {\bibfnamefont {F.}~\bibnamefont {Croccolo}}, \bibinfo {author} {\bibfnamefont {J.~M.}\ \bibnamefont {Ortiz~de Z{\'a}rate}},\ and\ \bibinfo {author} {\bibfnamefont {J.~V.}\ \bibnamefont {Sengers}},\ }\bibfield  {title} {\bibinfo {title} {Non-local fluctuation phenomena in liquids},\ }\href {https://doi.org/10.1140/epje/i2016-16125-3} {\bibfield  {journal} {\bibinfo  {journal} {Eur.\ Phys.\ J.\ E}\ }\textbf {\bibinfo {volume} {39}},\ \bibinfo {pages} {125} (\bibinfo {year} {2016})}\BibitemShut {NoStop}%
\bibitem [{\citenamefont {Vailati}\ \emph {et~al.}(2023)\citenamefont {Vailati}, \citenamefont {Bataller}, \citenamefont {Bou-Ali}, \citenamefont {Carpineti}, \citenamefont {Cerbino}, \citenamefont {Croccolo}, \citenamefont {Egelhaaf}, \citenamefont {Giavazzi}, \citenamefont {Giraudet}, \citenamefont {Guevara-Carrion}, \citenamefont {Horv{\'a}th}, \citenamefont {Köhler}, \citenamefont {Mialdun}, \citenamefont {Porter}, \citenamefont {Schwarzenberger}, \citenamefont {Shevtsova},\ and\ \citenamefont {Wit}}]{VailatiBatallerBouAliEtAl2023}%
  \BibitemOpen
  \bibfield  {author} {\bibinfo {author} {\bibfnamefont {A.}~\bibnamefont {Vailati}}, \bibinfo {author} {\bibfnamefont {H.}~\bibnamefont {Bataller}}, \bibinfo {author} {\bibfnamefont {M.~M.}\ \bibnamefont {Bou-Ali}}, \bibinfo {author} {\bibfnamefont {M.}~\bibnamefont {Carpineti}}, \bibinfo {author} {\bibfnamefont {R.}~\bibnamefont {Cerbino}}, \bibinfo {author} {\bibfnamefont {F.}~\bibnamefont {Croccolo}}, \bibinfo {author} {\bibfnamefont {S.~U.}\ \bibnamefont {Egelhaaf}}, \bibinfo {author} {\bibfnamefont {F.}~\bibnamefont {Giavazzi}}, \bibinfo {author} {\bibfnamefont {C.}~\bibnamefont {Giraudet}}, \bibinfo {author} {\bibfnamefont {G.}~\bibnamefont {Guevara-Carrion}}, \bibinfo {author} {\bibfnamefont {D.}~\bibnamefont {Horv{\'a}th}}, \bibinfo {author} {\bibfnamefont {W.}~\bibnamefont {Köhler}}, \bibinfo {author} {\bibfnamefont {A.}~\bibnamefont {Mialdun}}, \bibinfo {author} {\bibfnamefont {J.}~\bibnamefont {Porter}}, \bibinfo {author} {\bibfnamefont {K.}~\bibnamefont {Schwarzenberger}}, \bibinfo {author}
  {\bibfnamefont {V.}~\bibnamefont {Shevtsova}},\ and\ \bibinfo {author} {\bibfnamefont {A.~D.}\ \bibnamefont {Wit}},\ }\bibfield  {title} {\bibinfo {title} {Diffusion in liquid mixtures},\ }\href {https://doi.org/10.1038/s41526-022-00246-z} {\bibfield  {journal} {\bibinfo  {journal} {npj Microgravity}\ }\textbf {\bibinfo {volume} {9}},\ \bibinfo {pages} {1} (\bibinfo {year} {2023})}\BibitemShut {NoStop}%
\bibitem [{\citenamefont {Kim}\ \emph {et~al.}(2018)\citenamefont {Kim}, \citenamefont {Nonaka}, \citenamefont {Bell}, \citenamefont {Garcia},\ and\ \citenamefont {Donev}}]{KimNonakaBellGarciaDonev2018}%
  \BibitemOpen
  \bibfield  {author} {\bibinfo {author} {\bibfnamefont {C.}~\bibnamefont {Kim}}, \bibinfo {author} {\bibfnamefont {A.}~\bibnamefont {Nonaka}}, \bibinfo {author} {\bibfnamefont {J.~B.}\ \bibnamefont {Bell}}, \bibinfo {author} {\bibfnamefont {A.~L.}\ \bibnamefont {Garcia}},\ and\ \bibinfo {author} {\bibfnamefont {A.}~\bibnamefont {Donev}},\ }\bibfield  {title} {\bibinfo {title} {Fluctuating hydrodynamics of reactive liquid mixtures},\ }\href {https://doi.org/10.1063/1.5043428} {\bibfield  {journal} {\bibinfo  {journal} {J.\ Chem.\ Phys.}\ }\textbf {\bibinfo {volume} {149}},\ \bibinfo {pages} {084113} (\bibinfo {year} {2018})}\BibitemShut {NoStop}%
\bibitem [{\citenamefont {Jung}\ \emph {et~al.}(2026)\citenamefont {Jung}, \citenamefont {Kim}, \citenamefont {Garcia}, \citenamefont {Nonaka}, \citenamefont {Bell}, \citenamefont {Srivastava},\ and\ \citenamefont {Kim}}]{JungKimGarciaNonakaBellSrivastavaKim2026}%
  \BibitemOpen
  \bibfield  {author} {\bibinfo {author} {\bibfnamefont {H.~T.}\ \bibnamefont {Jung}}, \bibinfo {author} {\bibfnamefont {H.}~\bibnamefont {Kim}}, \bibinfo {author} {\bibfnamefont {A.~L.}\ \bibnamefont {Garcia}}, \bibinfo {author} {\bibfnamefont {A.~J.}\ \bibnamefont {Nonaka}}, \bibinfo {author} {\bibfnamefont {J.~B.}\ \bibnamefont {Bell}}, \bibinfo {author} {\bibfnamefont {I.}~\bibnamefont {Srivastava}},\ and\ \bibinfo {author} {\bibfnamefont {C.}~\bibnamefont {Kim}},\ }\bibfield  {title} {\bibinfo {title} {Thermodynamically consistent incorporation of the {L}angmuir adsorption model into compressible fluctuating hydrodynamics},\ }\href {https://doi.org/10.1063/5.0306932} {\bibfield  {journal} {\bibinfo  {journal} {J.\ Chem.\ Phys.}\ }\textbf {\bibinfo {volume} {164}},\ \bibinfo {pages} {094103} (\bibinfo {year} {2026})}\BibitemShut {NoStop}%
\bibitem [{\citenamefont {Srivastava}\ \emph {et~al.}(2023)\citenamefont {Srivastava}, \citenamefont {Ladiges}, \citenamefont {Nonaka}, \citenamefont {Garcia},\ and\ \citenamefont {Bell}}]{SrivastavaLadigesNonakaGarciaBell2023}%
  \BibitemOpen
  \bibfield  {author} {\bibinfo {author} {\bibfnamefont {I.}~\bibnamefont {Srivastava}}, \bibinfo {author} {\bibfnamefont {D.~R.}\ \bibnamefont {Ladiges}}, \bibinfo {author} {\bibfnamefont {A.~J.}\ \bibnamefont {Nonaka}}, \bibinfo {author} {\bibfnamefont {A.~L.}\ \bibnamefont {Garcia}},\ and\ \bibinfo {author} {\bibfnamefont {J.~B.}\ \bibnamefont {Bell}},\ }\bibfield  {title} {\bibinfo {title} {Staggered scheme for the compressible fluctuating hydrodynamics of multispecies fluid mixtures},\ }\href {https://doi.org/10.1103/PhysRevE.107.015305} {\bibfield  {journal} {\bibinfo  {journal} {Phys.\ Rev.\ E}\ }\textbf {\bibinfo {volume} {107}},\ \bibinfo {pages} {015305} (\bibinfo {year} {2023})}\BibitemShut {NoStop}%
\bibitem [{\citenamefont {Kim}\ \emph {et~al.}(2017)\citenamefont {Kim}, \citenamefont {Nonaka}, \citenamefont {Bell}, \citenamefont {Garcia},\ and\ \citenamefont {Donev}}]{KimNonakaBellGarciaDonev2017}%
  \BibitemOpen
  \bibfield  {author} {\bibinfo {author} {\bibfnamefont {C.}~\bibnamefont {Kim}}, \bibinfo {author} {\bibfnamefont {A.}~\bibnamefont {Nonaka}}, \bibinfo {author} {\bibfnamefont {J.~B.}\ \bibnamefont {Bell}}, \bibinfo {author} {\bibfnamefont {A.~L.}\ \bibnamefont {Garcia}},\ and\ \bibinfo {author} {\bibfnamefont {A.}~\bibnamefont {Donev}},\ }\bibfield  {title} {\bibinfo {title} {Stochastic simulation of reaction-diffusion systems: A fluctuating-hydrodynamics approach},\ }\href {https://doi.org/10.1063/1.4978775} {\bibfield  {journal} {\bibinfo  {journal} {J.\ Chem.\ Phys.}\ }\textbf {\bibinfo {volume} {146}},\ \bibinfo {pages} {124110} (\bibinfo {year} {2017})}\BibitemShut {NoStop}%
\bibitem [{SM()}]{SM}%
  \BibitemOpen
  \href@noop {} {}\bibinfo {note} {See Supplemental Material at [URL] for additional details on additional structure factor spectra and the derivation of an approximate structure factor for a hypothetical nonreactive system.}\BibitemShut {Stop}%
\bibitem [{\citenamefont {Ortiz~de Z{\'a}rate}\ and\ \citenamefont {Redondo}(2001)}]{ZarateRedondo2001}%
  \BibitemOpen
  \bibfield  {author} {\bibinfo {author} {\bibfnamefont {J.~M.}\ \bibnamefont {Ortiz~de Z{\'a}rate}}\ and\ \bibinfo {author} {\bibfnamefont {L.~M.}\ \bibnamefont {Redondo}},\ }\bibfield  {title} {\bibinfo {title} {Finite-size effects with rigid boundaries on nonequilibrium fluctuations in a liquid},\ }\href {https://doi.org/10.1007/s100510170223} {\bibfield  {journal} {\bibinfo  {journal} {Eur.\ Phys.\ J.\ B}\ }\textbf {\bibinfo {volume} {21}},\ \bibinfo {pages} {135} (\bibinfo {year} {2001})}\BibitemShut {NoStop}%
\bibitem [{FHD()}]{FHDeX}%
  \BibitemOpen
  \href {https://github.com/AMReX-FHD/FHDeX.git} {\bibinfo {title} {{FHDeX, https://github.com/AMReX-FHD/FHDeX.git}}}\BibitemShut {NoStop}%
\end{thebibliography}
